\documentclass[nologo,11pt,a4paper]{ETHpaper}
\usepackage{verbatim}  
\usepackage{graphicx}                  
\usepackage{subfigure}
\usepackage{setspace}

\usepackage[sort&compress]{natbib}     
\usepackage{dcolumn}
\usepackage{amssymb}
\usepackage{amsfonts}
\usepackage{array}
\usepackage{amsthm}
\usepackage{amsmath}
\usepackage{caption}
\usepackage{url}
\usepackage{appendix}

\usepackage{color}

\title{\textbf{The Rise and Fall of R\&D Networks}}

\titlealternative{The Rise and Fall of R\&D Networks\\
\textit{ICC -- Industrial and Corporate Change} (2016); doi: 10.1093/icc/dtw041}

\author{Mario V. Tomasello,$^{*,1}$  Mauro Napoletano,$^{2,3}$  Antonios Garas$^1$ and Frank Schweitzer$^1$}

\address{
$^*$ Corresponding author; E-mail: mtomasello@ethz.ch\\
$^1$ Chair of Systems Design, Department of Management, Technology and Economics,\\ ETH Zurich, Weinbergstrasse 56/58, 8092 Zurich, Switzerland.\\
$^2$ OFCE -- SKEMA Business School, Campus de Sophia Antipolis 60, rue Dosto\"ievski\\ BP 85 06902 Sophia Antipolis Cedex, France.\\
$^3$ Sant'Anna School of Advanced Studies, Piazza Martiri della Libert\`{a} 33,\\ 56127 Pisa, Italy. 
}

\authoralternative{M. V. Tomasello,  M. Napoletano,  A. Garas, F. Schweitzer}

\date{}

\begin{document}

\maketitle

\begin{abstract}
  \noindent Drawing on a large database of publicly announced R\&D
  alliances, we empirically investigate the evolution of R\&D networks and the process
  of alliance formation in several
  manufacturing sectors over a 24-year  period (1986-2009). Our goal
  is to empirically evaluate the temporal and sectoral robustness of a
  large set of network indicators, thus providing a more complete
  description of R\&D networks with respect to the existing
  literature.  We find that most network properties are not only
  invariant across sectors, but also independent of the scale of
  aggregation at which they are observed, and we highlight
the presence of core-periphery architectures in explaining some
properties emphasized in previous empirical studies (e.g. asymmetric
degree distributions and small worlds).  In addition, we show that many properties of R\&D networks are
  characterized by a rise-and-fall dynamics with a peak in the
  mid-nineties. We find that such dynamics is driven by
  mechanisms of accumulative advantage, structural homophily and
  multiconnectivity. In particular, the change from the ``rise'' to the ``fall''
  phase is associated to a structural break in the importance of multiconnectivity.
\end{abstract}

\begin{spacing}{1.2}
  \setlength{\tabcolsep}{3pt} \renewcommand{\arraystretch}{1.0}

  \section{Introduction}
  This work investigates the structural properties of empirical R\&D
  networks and the rules of alliance formation by firms. In several
  industries, and especially in those with rapid technological growth,
  innovation relies on general and abstract knowledge often built on
  scientific research \citep{powell1996interorganizational,
    Dosi82}. This has allowed for a division of innovative labor and
  fostered collaboration across firms
  \citep{arora1994changing,arora1994evaluating,
    dosi95:_learn_market_selec_evolut_indus_struc}. Accordingly, the
  last three decades have witnessed a significant growth in the number
  of formal and informal R\&D collaborations
  \citep[e.g.][]{hagedoorn2002inter,powell2005network}, and several
  studies have documented the importance of networks for knowledge
  spillovers and firms' innovative performance \citep[see
  e.g.][]{powell1996interorganizational,ahuja2000collaboration,GiulianiJEG}

  The growing importance of R\&D networks has resulted in a
  signficant amount of empirical research about the structural
  properties of those networks and on the determinants of their
  evolution. On the one hand, these empirical works have shown that R\&D networks are typically
  sparse and characterized by heavily asymmetric distributions of the
  number of alliances
  \citep[e.g.][]{powell2005network, rosenkopf2007comparing,
    hanaki2010dynamics}. Furthermore, R\&D networks exhibit the
  so-called small world property \citep[as shown
  by][]{fleming06:_manag_creat_in_small_world,
    fleming07:_small_world_and_region_innov}, i.e. they are
  characterized by short average path length and high clustering
  \citep{strogatz98:_collective_dynamics_small_world_networks}. 
On the other hand, another group of empirical studies \citep[see e.g.][]{gulati2012:_rise_and_fall_of_small_world,powell2005network,rosenkopf2008investigating,gulati1995social} has proposed and tested models for the
process of alliance formation driving the evolution of R\&D networks.
This latter research stream is firmly rooted on the idea that the
process of network evolution is strongly path-dependent.
In that, the existing structures of the network, and the position of the firms therein,
capture different technological
as well as social and organizational characteristics, and shape firms'
decisions about future creation and deletion of alliances.\footnote{Besides,
the hypothesis about the path-dependent character of R\&D network
evolution also underlies another stream of theoretical works, which
in the latter years have tried to account for the observed properties
of R\&D networks and their dynamics \citep[see in particular the works
of][]{goyal2003networks,konig2012efficiency,koenig2013nestedness,cowan2004network}.}

The above empirical studies have greatly contributed to the understanding of empirically
  observed R\&D networks. However, they have often focused only on a
  small number of industries and/or they have rarely considered how
  the properties of the network may evolve over time. Finally, they have
  focused on a small set of network measures (e.g. size, degree
  heterogeneity, small world properties), which limits the
  understanding of the path-dependent process of alliance formation.

  On these premises, our work improves on the foregoing literature
  along several dimensions. \textit{First}, we analyze a global
  inter-firm R\&D network (the \textit{pooled} R\&D network), as well
  as its decomposition in a series of subnetworks for several
  representative manufacturing sectors (the \textit{sectoral} R\&D
  networks).  Through such an analysis, we are able to check whether
  the network properties that have been analyzed by the current
  literature for sectors like computers
  \citep[e.g.][]{hanaki2010dynamics} or pharmaceuticals
  \citep[e.g.][]{powell2005network} are robust across different
  sectors of activity. In addition, by comparing the properties at the
  pooled and at the sectoral levels, we are able to check for the
  presence of \textit{universal} properties of R\&D networks that hold
  irrespectively of the scale of aggregation at which they are
  observed.

 \textit{Second}, we investigate a broad set of network properties. The
  object of our analysis are not only the basic measures that have so far been
  considered in the empirical literature (size, degree heterogeneity,
  small world properties), but also indicators related to more complex
  features of the network, such as assortativity \citep[i.e. the
  presence of positive correlation in the number of alliances among
  firms; see also][]{newman03:_mixin_patter_in_networ} 
  and the presence of ``nested'' core-periphery architectures
  \citep[see][]{bascompte2003nestedness}. 
  In this way, we refine the existing knowledge on R\&D networks by detecting
  new stylized facts about the structural features of those
  networks and shed further lights on the drivers of the process of
  alliance formation.

  \textit{Third}, building on the above-mentioned structural analysis,
  we perform a longitudinal analysis of the
  determinants of R\&D alliance formation. In this analysis, our dependent variable
  is a firm dyad, and the observation unit is every
  potential pair of firms in the R\&D network. We then investgate
  which combination of attachment rules provides a good description of the empirical
  evolution of alliances in the sample considered. We focus on
  the mechanisms of alliance formation which have received more
  attention in the literature so far, namely:  (i)
  \textit{accumulative advantage}, (ii) \textit{structural homophily}
  (or \textit{diversity}) and (iii) \textit{multiconnectivity}
  \citep[see][]{rosenkopf2008investigating, powell2005network}.
  Moreover, we conduct regression analyses for different time periods.
  In this way we are able to check if different attachment rules may
  account for different evolutions of the network over the observed
  sample. Finally, we also consider separately alliances formed among incumbent
  firms in the network, and alliances where at least one firm is
  entering the network, in order to check whether drivers of alliance
  formation are different across incumbents and entrants.

Our results show, first, that the evolution of R\&D networks has been
universal across different scales of aggregation. Indeed, many structural
properties of the network (e.g. asymmetric degree distributions,
assortativity, presence of small worlds) robustly hold both when alliances
are considered irrespectively of the sectors of the firms and when
sectoral networks are analyzed. Second, we show that the dynamics of
R\&D networks has been characterized by two distinctive phases: in the
first of them (the ``rise'' phase, from 1986 to 1997), alliances gave rise to dense
network structures, organized into very few large components
displaying core-periphery nested architectures. In the
second phase (the ``fall'' phase, from 1998 to 2009), networks become
more sparse and fragmented into many small components with few firms. 
Third, our regression results bring support to the idea that the process of
alliance formation has largely been driven by accumulative advantage
and by the search for similar partners (``homophily''). Moreover, we find
that in the rise phase firms also tried to form alliances to
increase the number of paths through which other firms could be
directly or indirectly reached (``multiconnectivity''), whereas in the
fall phase such an alliance driver lost significance. In turn, such a
structural break in the importance of multiconnectivity underlies 
the emergence of densely connected networks and their subsequent
fragmentation. 

Our results have several implications. First, the universality of the
structural properties of the network reinforces the idea that R\&D alliances can be analyzed independently of the
characteristics of the sectors to which firms belong, and by focusing
on simple rules of alliance formation capturing different
organizational and technological drivers, given the
existing network structure. Second, our findings on the rise and fall
of the networks confirm previous results about the presence of a
life-cycle in the evolution of networks \citep[see in
particular][]{gulati2012:_rise_and_fall_of_small_world}. At the same
time, they show that such a dynamics is not specific to a single
industry but is rather a general property of many sectoral networks
and it also holds independently of the scale of aggregation. In
addition, the finding that network components are organized into core-periphery
architectures in the rise phase is able to jointly explain two
features that have so far received a great deal of attention in the empirical
literature, namely the presence of small worlds and
fat-tailed degree distributions. Finally,
and in line with \citet{powell2005network}, our results show that
multiconnectivity matters besides more traditional drivers of R\&D
alliances, and in particular to explain structural breaks in the R\&D network evolution.

The paper is organized as follows. Section \ref{sec:data} describes
the data and the methodology used to build the networks of R\&D
alliances. Section \ref{sec:basic-netw-prop} presents a set of basic
network properties, such as size, density, the emergence of a giant
component, and discusses their evolution over time. Section
\ref{sec:homoph-hetero-alli-behav} studies the heterogeneity and the
homophily in the network, by analyzing degree distributions and degree
correlations (assortativity).  Section \ref{sec:small-worlds-comm}
studies the emergence of small-world and nested core-periphery structures in the R\&D
networks. Section \ref{sec:alliance_formation} investigates the
determinants of alliance formation through a set of regression models,
discussing the results in light of the existing theoretical and
empirical literature on R\&D networks. Finally, Section
\ref{sec:concluding-remarks} concludes. The Appendix contains
a description of all network measures used in the paper.

  \section{R\&D alliance data and the construction of R\&D networks}
  \label{sec:data}

  An \textit{R\&D network} is a representation of the research and
  development alliances occurring between firms in one or more
  industrial sectors within a given period. Every network consists
  of a set of \textit{nodes} and \textit{links} connecting pairs of
  nodes. In our representation, each node of the network is a firm and
  every link represents an R\&D alliance between two firms. By R\&D
  alliance, we refer to an event of partnership between two firms,
  that can span from formal joint ventures to more informal research
  agreements, specifically aimed at research and development
  purposes. To detect such events, we use the \textit{SDC Platinum}
  database, provided by Thomson Reuters, that reports all publicly
  announced alliances, from 1984 to 2009, between several kinds of
  economic actors (including manufacturing firms, investors, banks and
  universities). We then select all the alliances concerning
  manufacturing firms and displaying the 
  ``R\&D'' tag; after applying this filter, we obtain a total of
  8,835 listed alliances.

  Information in the SDC dataset is gathered only from announcements
  in public sources, such as press releases or journal
  articles. Nevertheless, despite the bias that could be introduced by
  such a collection procedure,
  \citet{schilling2009understanding_alliance_data} shows that the SDC
  Thomson dataset provides a consistent picture with respect to
  alternative databases \citep[e.g. CORE and MERIT-CATI, see also][]{hagedoorn00:_resear} in terms of
  alliance activity over time, industry composition and geographical
  location of companies.  The country coverage of the
  SDC dataset is also consistent with the alternative datasets: 55\% of the listed firms are registered in the
  U.S.A., 8.5\% in Japan, 4.4\% in Canada, 4.3\% in the U.K. and so
  on. See Table \ref{table:geographical_composition} for more details.

\begin{table}[h!]
  \scriptsize
  \centering
  \begin{tabular*}{0.45\linewidth}{@{\extracolsep{\fill}}l*{2}{@{ }r}}
    \hline
    & \textbf{Number of firms} & \textbf{Fraction} \cr
    \hline

    \textbf{Pooled}    & 9499 & 1.000 \cr 
    United States      & 5245 & 0.552 \cr 
    Japan              &  804 & 0.085 \cr 
    Canada             &  421 & 0.044 \cr 
    United Kingdom     &  411 & 0.043 \cr 
    Germany            &  358 & 0.038 \cr 
    China              &  331 & 0.035 \cr 
    France             &  236 & 0.025 \cr 
    Australia          &  202 & 0.021 \cr 
    India              &  119 & 0.013 \cr 
    Italy              &  109 & 0.011 \cr 
    \textbf{Other}     & 1263 & 0.133 \cr 

    \hline \hline

  \end{tabular*}
  \caption{Network composition with respect to the geographical provenience of the firms listed in our dataset. The top-ten represented countries account for 87\% of all listed firms.}
  
  \label{table:geographical_composition}

\end{table}

We check all firm names and control for all legal extensions
(e.g. ``ltd'', ``inc'', etc.) and other recurrent keywords
(e.g. ``bio'', ``tech'', ``pharma'', ``lab'', etc.) that could affect
the matching between entries referring to the same firm. We keep as
separated entities the subsidiaries of the same firm located in
different countries. The raw dataset contains a total of 16,313 firms,
which are reduced to 9,499 after running such an extensive
standardization procedure.

In our network representation, we draw a link connecting two nodes
every time an alliance between the two corresponding firms is
announced in the dataset. An alliance is associated with an undirected
link, as we do not have any information about the initiator of the
alliance. When an alliance involves more than two firms
(\textit{consortium}), all the involved firms are connected in pairs,
resulting into a fully connected clique.  Following this procedure,
the 8,835 alliance events listed in the dataset result in a total of
11,827 links.  Similarly to \citet{rosenkopf2007comparing}, the R\&D
network that we consider in our study is \textit{unipartite}, as we only
have one set of actors (``the firms''), whose elements may be
connected -- or not -- by publicly announced alliances.\footnote{Our
  work differs from previous empirical studies
  \citep[e.g.][]{lissoni2013smallworlds,hanaki2010dynamics,cantner2006network}
  which construct the network through the association of firms with
  patents and/or inventors. Those studies use patent data to build the
  network and associate elements in the set ``firms'' to the elements
  in the set ``patents''. This way, the network they obtain is
  \textit{bipartite}.}

Multiple links between the same nodes are in principle allowed (two
firms can have more than one alliance on different
projects). Nevertheless, as we aim at studying the connections between
firms, and not the number of alliances a firm is involved in, we
discard this information and use unweighted links in our network
representation. For this reason, we define the degree of a node as the
number of other nodes to which it is linked, i.e. the number of
partners that a firm has -- not the number of alliances. Furthermore,
a firm appears in the R\&D network only if it is involved in at least
one alliance. Our study is focused exclusively on the embeddedness of
firms into an alliance network; for this reason, isolated nodes are
not part of our network representation.

Both the links and the nodes of the R\&D network are characterized by
an entry/exit dynamics. Alliances between firms have a finite duration
\citep[see][]{deeds1999examination,phelps2003technological}. This
causes some firms to disappear from the network, after they no longer
participate in any alliance. Likewise, many new firms that are not
listed in any previous alliance may enter the network at the beginning
of a new year.  Our longitudinal study obviously requires precise
temporal information about the formation and the deletion of
alliances. The SDC Platinum dataset contains the beginning date of
every alliance, but there is no information about any of the ending
dates (firms do not usually organize press releases to announce the
end of an alliance). We are thus forced to make some assumptions about
the alliance durations. We draw the duration of every
alliance from a normal distribution with mean value from 1 to 5 years
and standard deviation from 1 to 5 years, and we find that all our
results remain qualitatively unchanged 
within these ranges.\footnote{The
variation of the standard deviation has nearly no influence on the
patterns exhibited by all network measures, whereas a variation of the mean would only shift some trends in terms of absolute value, but not in terms of time-evolution and peak positions.}
Given the strong robustness of the R\&D network to
the variation of alliance lengths, we take a conservative approach and
assume a fixed 3-year length for every partnership, consistently with
previous empirical work
\citep[e.g.][]{rosenkopf2007comparing,phelps2003technological,deeds1999examination}.
More precisely, we link two nodes when an alliance between the
corresponding firms occurs and we delete this link 3 years after its
formation. This is also the reason why, even though our dataset starts reporting 
R\&D alliances from the year 1984, we start building the corresponding R\&D network from the year 1986.
In this way, we are able to build 26 snapshots of the R\&D
network -- one for every year -- from 1986 to 2009.  From now on we
call the network containing all companies, irrespectively of their
industrial sector, the \textit{pooled R\&D network}.

Every firm in the dataset is associated to a SIC code (Standard
Industrial Classification).  This allows us to build a series of
\textit{sectoral R\&D networks}, one for each sector that we identify
in the dataset.  A sectoral R\&D network contains only alliances in which at least \textit{one} of the
partners has a three-digit SIC code matching the selected sector
\citep[see also][for a similar approach]{rosenkopf2007comparing}. The
rules for link deletion are the same as in the pooled R\&D network.
More precisely, we focus on the 10 largest manufacturing 
sectors (in terms of numbers of firms in the network). Table
\ref{table:sectoral_composition} provides the list of the sectors
considered, together with the number of reported firms
and alliances, both in absolute and in relative terms.

\begin{table}[h!]
  \scriptsize
  \centering
  \begin{tabular*}{0.82\linewidth}{@{\extracolsep{\fill}}l*{2}{@{ }r}|*{2}{@{ }r}}
    \hline
    & \textbf{Number of firms} & \textbf{(fraction)} & \textbf{Number of alliances} & \textbf{(fraction)} \cr
    \hline

    \textbf{Pooled} & 9499 & 1.000 & 8835 & 1.000 \cr 
    Pharmaceuticals & 2224 & 0.234 & 2576 & 0.292 \cr 
    Computer Software & 1826 & 0.192 & 1533 & 0.174 \cr 
    Electronic Components &  596 & 0.063 &  787 & 0.089 \cr 
    Computer Hardware &  498 & 0.052 &  741 & 0.084 \cr 
    Medical Supplies &  439 & 0.046 &  285 & 0.032 \cr 
    Communications Equipment &  436 & 0.046 &  399 & 0.045 \cr 
    Laboratory Apparatus &  301 & 0.032 &  207 & 0.023 \cr 
    Motor Vehicles &  243 & 0.026 &  240 & 0.027 \cr 
    Inorganic Chemicals &  147 & 0.015 &  129 & 0.015 \cr 
    Aircrafts and parts &  146 & 0.015 &  132 & 0.015 \cr 
    \textbf{Other} & 2643 & 0.278 & 1805 & 0.204 \cr 

    \hline \hline
  \end{tabular*}
  \caption{Network composition with respect to the industrial sectors. The top-ten represented sectors account for 72\% of all listed firms and 80\% of all listed alliances. \textit{Note:} when an alliance event involves firms from different sectors, the weight of the alliance is equally distributed between the partners.}  
  \label{table:sectoral_composition}
\end{table}

We study both the pooled R\&D network and the sectoral R\&D networks
by computing a set of network indicators along the whole observation
period.
We group our descriptive analysis of the network into three sections. We begin by discussing
basic facts about the evolution of the network, such as its size,
density and connectedness. Next, we discuss the degree of
heterogeneity and homophily in the network, by studying the evolution
of the degree distributions and of assortativity patterns. Finally, we
discuss how network components are organized, by studying the presence of
small worlds and of core-periphery structures.

\section{Basic facts about the evolution of R\&D networks}
\label{sec:basic-netw-prop}

\begin{figure}[h!]
  \vspace{8pt}
  \begin{center}
    \includegraphics[width=1.\textwidth,angle=0]{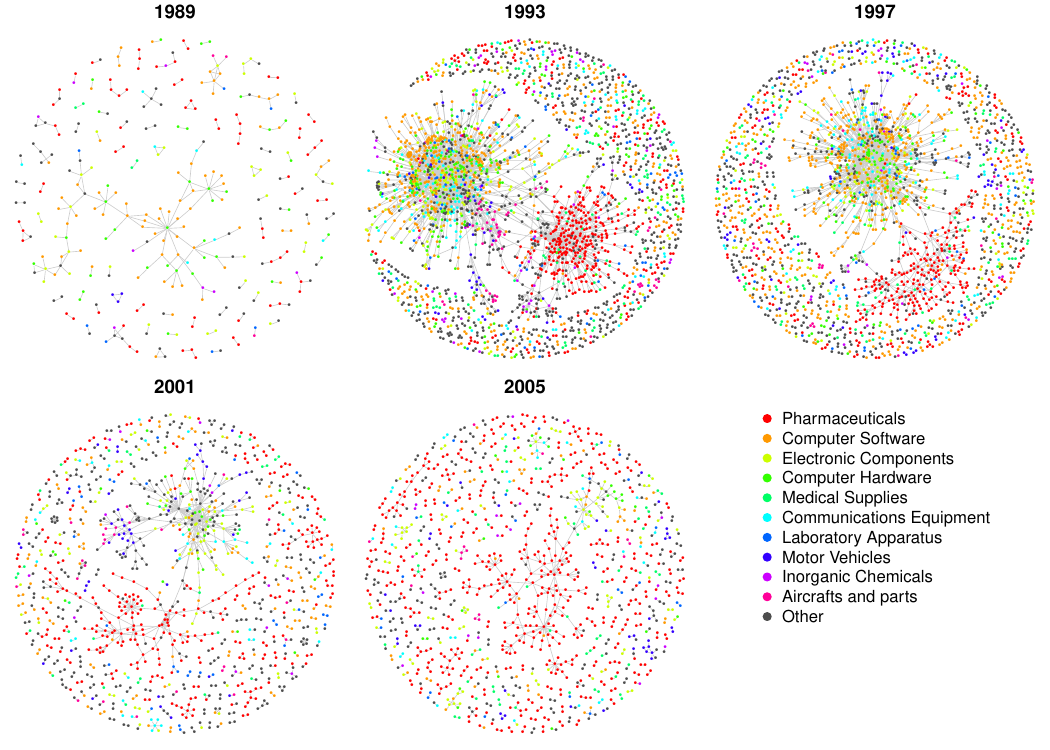}
  \end{center}
  \vspace{-8pt}
  \caption[]{\textbf{Evolution of the pooled R\&D network.} Pooled
    R\&D network snapshots in 1989, 1993, 1997, 2001 and 2005. In
    order to ease the visualization, we only plot the nodes belonging
    to the ten largest sectors and their alliance partners.}
  \label{fig:aggr_net}
\end{figure}

\begin{figure}[h!]
  \vspace{8pt}
  \begin{center}
    \includegraphics[width=1.\textwidth,angle=0]{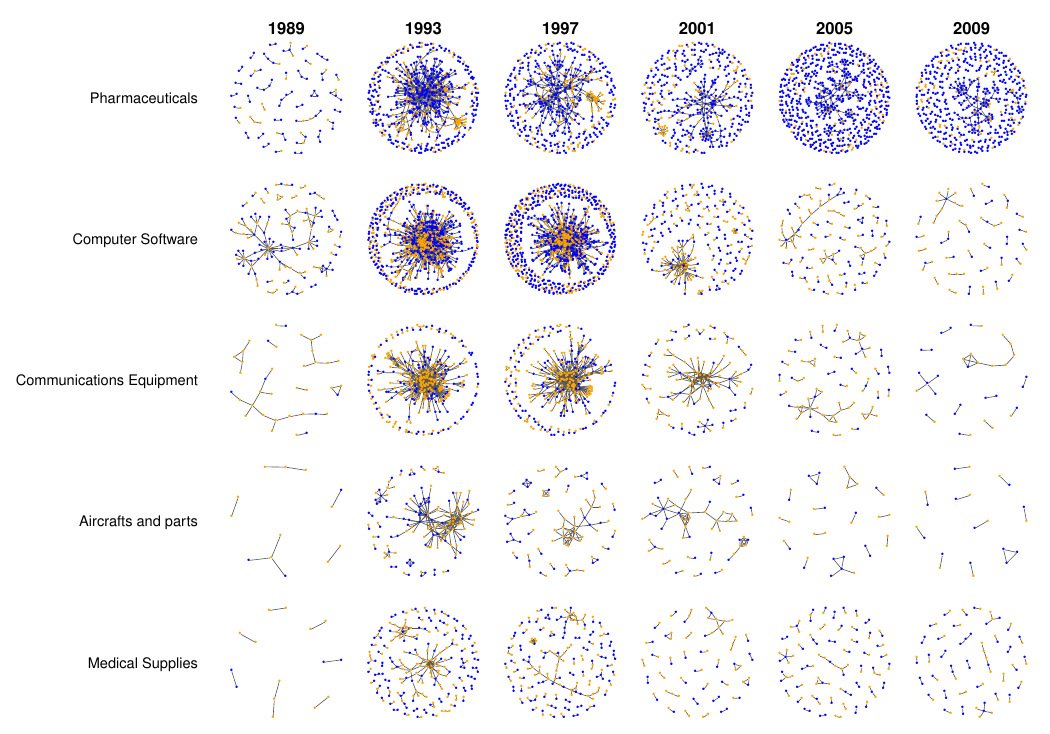}
  \end{center}
  \vspace{-8pt}
  \caption[]{\textbf{Evolution of five selected sectoral R\&D
      networks.} Snapshots in 1989, 1993, 1997, 2001, 2005 and 2009
    for five selected sectoral R\&D networks: Pharmaceuticals,
    Computer Software, Communication Equipments, Aircrafts and parts,
    Medical Supplies.  Blue nodes represent the firms strictly
    belonging to the examined sector, while orange nodes represent
    their alliance partners belonging to different sectors.}
  \label{fig:sec_net}
\end{figure}

We begin our analysis by discussing some basic properties concerning
the evolution of the pooled R\&D network. Following
\citet{powell2005network,rosenkopf2007comparing,rosenkopf2008investigating}
we employ network visualization techniques to provide a first
assessment of how network structures evolved over the years
analyzed. More precisely, Figures \ref{fig:aggr_net} and
\ref{fig:sec_net} show several snapshots of, respectively, the pooled and
five sectoral networks.  The plots are produced using the
\textit{igraph} library\footnote{The \textit{igraph} library is freely
  available at \url{http://igraph.sourceforge.net/}.} for \textit{R},
and the networks are displayed using the Fruchterman-Reingold
algorithm \citep[cf.][]{fruchterman_reingold_1991}. This is a
force-based algorithm for network visualization which positions the
nodes of a graph in a two-dimensional space so that all the edges are
of similar length and there are as few crossing edges as possible. The
result is that the most interconnected nodes are displayed close to
each other in the resulting two-dimensional plot. We use node colors
to identify the sectors to which firms belong. More precisely, in
Figure \ref{fig:aggr_net} each different color indicates a different
sector. In Figure \ref{fig:sec_net}, instead,
different colors indicate whether the firm belongs to the same sector
on which the network is centered or not. This is in order to provide a
visual indication on the share of intra-sectoral and inter-sectoral
alliances in each industry.

Figure \ref{fig:aggr_net} denotes the presence of different phases in
the evolution of the R\&D network. The plots suggest a
significant growth of the network until 1997, and a reversal of this
trend aftermath. Interestingly, such a
rise-and-fall pattern is also present in sectoral networks. Indeed,
Figure \ref{fig:sec_net} shows that -- although with different
intensities -- all plotted sectors display a concentration of
alliance activities in 1993 and 1997, followed by a decline in the
number of alliances in the 2000s decade. Incidentally, notice that the same rise-and-fall dynamics is displayed by sectors which are very
different in terms of technological characteristics
\citep[e.g. Pharmaceutical and Aircrafts and Parts,
see][]{rosenkopf2007comparing}.

\begin{figure}[h!]
  \begin{center}
    \includegraphics[width=0.55\textwidth,angle=0]{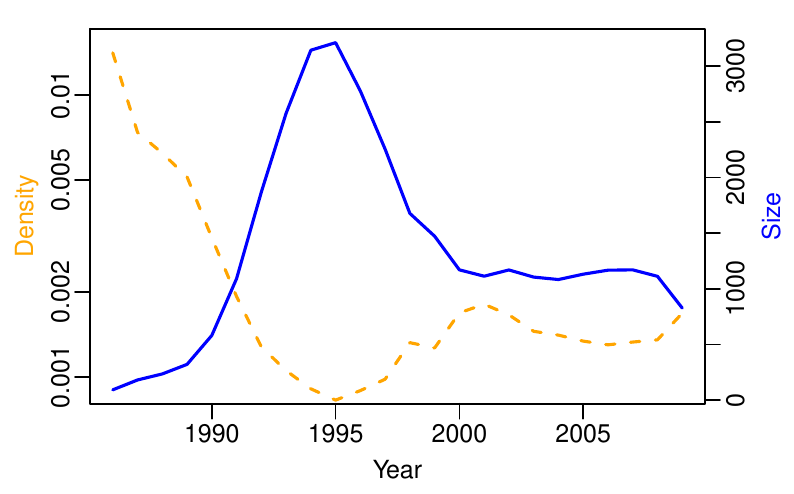}
  \end{center}
  \vspace{-8pt}
  \caption[]{\textbf{Size and density evolution of the pooled R\&D
      network.} Time-evolution of size (solid line, right y-axis) and
    density (dashed line, left y-axis) of the pooled R\&D network.
    Note that the density is visualized on a logarithmic scale.}
  \label{fig:density}
\end{figure}

Figure \ref{fig:density} provides important additional elements about the
network dynamics in our sample. The figure shows the evolution of the network density
(number of existing links divided by the number of all possible links
in the network) and the network size of the pooled network, and shows quite starkly that the
growth in the size of the network has been associated to a significant
fall in its density.  This means that the expansion of the R\&D
network was heavily driven by new alliances created by entrant
firms. Moreover, after the ``golden age'', the fall of the network has been 
associated with a decrease in the number of nodes. Because of the
importance of entry in the observed R\&D network dynamics, in Section \ref{sec:alliance_formation}
we perform separate regression analyses to investigate the
determinants of the formation of alliances by entrants.

Another interesting feature of network dynamics in our sample is the emergence of densely connected \textit{giant
  components}, both at the pooled and sectoral level. This is evident not only
from the plots in Figures \ref{fig:aggr_net} and
\ref{fig:sec_net}, but also from the time-evolution of the number of
firms in such components, and reported in Table
\ref{table:giant_comp}.\footnote{A connected component is defined as a set of
  nodes which are connected to each other by at least one path (i.e. a
  sequence of links). We refer to the largest connected component as
  the \textit{giant component} of the network. The giant component
  size to the overall network size ratio (or \textit{giant component
    fraction}) is a rough indicator of the network connectedness.} The
emergence of a giant component in the network is of particular
interest, as both previous empirical and theoretical works
\citep[e.g.][]{powell2005network,goyal2003networks,konig2012efficiency}
have stressed that high network connectedness favors
technological spillovers and overall knowledge growth by increasing
the number of knowledge sources to which single firms have direct
or indirect access via alliances. Figure \ref{fig:aggr_net}
also shows significant heterogeneity in terms of the type of sectors
present in the giant component although two categories of sectors
seems to be prevalent in the component: pharmaceuticals and
\textit{ICT}-related sectors (computer software and hardware,
electronic components, communications equipment). The foregoing giant
component has then significantly shrunk in the 2000s, leaving space to
a growing periphery of disconnected dyads (pairs of allied
firms). Such a process of increasing connectedness and
subsequent fragmentation of the network is present also at the
sectoral level (see Figure \ref{fig:sec_net}), although the intensity
of the network fragmentation in the 2000s looks much lower in
pharmaceuticals than in the other plotted sectors.

\begin{table}[h!]
  \scriptsize
  \centering
  \begin{tabular*}{0.99\linewidth}{@{\extracolsep{\fill}}l*{6}{@{ }r}}
    \hline
    & \textbf{1986-1989} & \textbf{1990-1993} & \textbf{1994-1997} & \textbf{1998-2001} & \textbf{2002-2005} & \textbf{2006-2009} \cr
    \hline

    \textbf{Pooled} & 207 & 1529 & 2846 & 1358 & 1122 & 1069 \cr 
    Pharmaceuticals &  63 &  440 &  617 &  461 &  508 &  666 \cr 
    Computer Software &  74 &  485 & 1145 &  406 &  198 &   94 \cr 
    Electronic Components &  66 &  312 &  528 &  246 &  188 &  144 \cr 
    Computer Hardware &  65 &  372 &  650 &  208 &   90 &   40 \cr 
    Medical Supplies &  10 &  142 &  236 &  120 &   86 &  111 \cr 
    Communications Equipment &  26 &  210 &  408 &  172 &  104 &   51 \cr 
    Laboratory Apparatus &  22 &  148 &  225 &  118 &   84 &   65 \cr 
    Motor Vehicles &  15 &  104 &  178 &  104 &   89 &   69 \cr 
    Inorganic Chemicals &  18 &  100 &  132 &   54 &   41 &   32 \cr 
    Aircrafts and parts &  12 &   82 &  127 &   65 &   42 &   24 \cr

    \hline \hline
  \end{tabular*}
  \caption{Network size for the pooled and the sectoral R\&D networks. The values are averages
    within each sub-period.}
  
  \label{table:size}

\end{table}

\begin{table}[htbp]
  \scriptsize
  \centering
  \begin{tabular*}{0.99\linewidth}{@{\extracolsep{\fill}}l*{6}{@{ }r}}
    
    \hline
    & \textbf{1986-1989} & \textbf{1990-1993} & \textbf{1994-1997} & \textbf{1998-2001} & \textbf{2002-2005} & \textbf{2006-2009} \cr
    \hline

    \textbf{Pooled} & 0.112 & 0.507 & 0.536 & 0.367 & 0.179 & 0.216 \cr 
    Pharmaceuticals & 0.107 & 0.583 & 0.663 & 0.495 & 0.240 & 0.334 \cr 
    Computer Software & 0.287 & 0.664 & 0.613 & 0.354 & 0.244 & 0.096 \cr 
    Electronic Components & 0.152 & 0.665 & 0.726 & 0.583 & 0.387 & 0.106 \cr 
    Computer Hardware & 0.257 & 0.681 & 0.757 & 0.652 & 0.526 & 0.152 \cr 
    Medical Supplies & 0.254 & 0.189 & 0.294 & 0.117 & 0.091 & 0.061 \cr 
    Communications Equipment & 0.222 & 0.665 & 0.697 & 0.577 & 0.400 & 0.181 \cr 
    Laboratory Apparatus & 0.242 & 0.439 & 0.372 & 0.311 & 0.181 & 0.088 \cr 
    Motor Vehicles & 0.289 & 0.537 & 0.529 & 0.472 & 0.311 & 0.098 \cr 
    Inorganic Chemicals & 0.253 & 0.373 & 0.408 & 0.147 & 0.231 & 0.266 \cr 
    Aircrafts and parts & 0.483 & 0.628 & 0.463 & 0.355 & 0.237 & 0.152 \cr

    \hline  \hline
  \end{tabular*}
  \caption{Fraction of the giant component for the pooled and the
    sectoral R\&D networks. The values are
    averages within each sub-periods.}
 
  \label{table:giant_comp}
\end{table}
The above analysis shows the existence of patterns that are
invariant to the scale of aggregation or the sector where they are
observed. Namely, both the pooled and sectoral R\&D networks
experience a robust growth in both size and connectedness until
1997. In particular, the years between 1994 and 1997 (the ``golden
age'' of R\&D networks), witness not only a higher number of
alliances, but also the emergence of a significantly large giant
component. This robust growth is then replaced by a decline phase,
characterized by both a reduction in the number of alliances and by the
breakdown of the network into smaller components. In the next
section, we add further details to the above picture of network
evolution by investigating heterogeneity and homophily in the 
formation of alliances.

\section{Heterogeneity and homophily in R\&D alliances}
\label{sec:homoph-hetero-alli-behav}

A good deal of literature has analyzed the properties of the degree
distributions in R\&D networks. Empirical studies have shown that
degree distributions in R\&D networks tend to be broad and highly
skewed. However, some studies find exponential distributions
\citep{riccaboni02:_firm_growt_in_networ}, while others find power-law
distributions \citep{powell2005network}. The presence of a power-law
distribution would indicate the existence of an underlying
multiplicative growth process
\citep{simon55:_class_of_skew_distr_funct,
  reed01:_paret_zipf_and_other_power_laws}. In the context of R\&D
networks the presence of skewed and fat-tailed distributions, such as power-laws, in the firms' degrees indicates
that a few firms have a disproportionate number of
ties compared to other firms. This may in turn indicate some form of \textit{accumulative advantage} at work in the process of
alliance formation, where firms that are able
to get an initial advantage position in a technological field, or in
terms of alliance experience, are then able to attract a large number
of partners \citep[see][]{powell2005network}. One model capturing the idea of accumulative advantage
is the so-called ``preferential attachment'' model by
\citet{barabasi99:_emerg}, which predicts the emergence of a power-law
degree distribution on the basis of a mechanism where entrant firms tend
to attach to incumbent partners with higher degree. Power-law
distributions can also emerge in network models
\citep[e.g.][]{koenig2013nestedness} where accumulative advantage is
captured by the centrality of the position of a firm in the
network. In this respect, connections to a more
central actor in the network can allow for higher knowledge growth, by
granting access to larger and more diversified knowledge sources
\citep[see][]{ahuja2000collaboration,powell2005network,konig2012efficiency}.

We contribute to the existing debate about degree
distributions in R\&D networks by studying their evolution over time
and comparing the results across different sectors.\footnote{As
  already mentioned in Section \ref{sec:data}, we define the degree as
  the number of partners of a firm, and not the number of
  alliances. In addition, we
  utilize the complementary cumulative distribution function (see
  Appendix) in order to display all the analyzed degree distributions,
  given its higher stability and ease of visualization.}
All the analyses performed in this Section consider networks which are obtained by pooling, i.e. adding up, all the observations for each node over a 4-year period.\footnote{We adopt this approach because it is the most suitable for the visualization of individual firm properties and their corresponding distributions, as opposed to global network measures (see Sections \ref{sec:small-worlds-comm}), which have to be computed separately in every year and then averaged.
However, we have found that our results for the heterogeneity and homophily indicators are not affected by double counting and are robust to the choice of averaging or pooling the observations over 4-year time periods.
}

\begin{figure}[h!]
  \begin{center}
    \includegraphics[width=1\textwidth,angle=0]{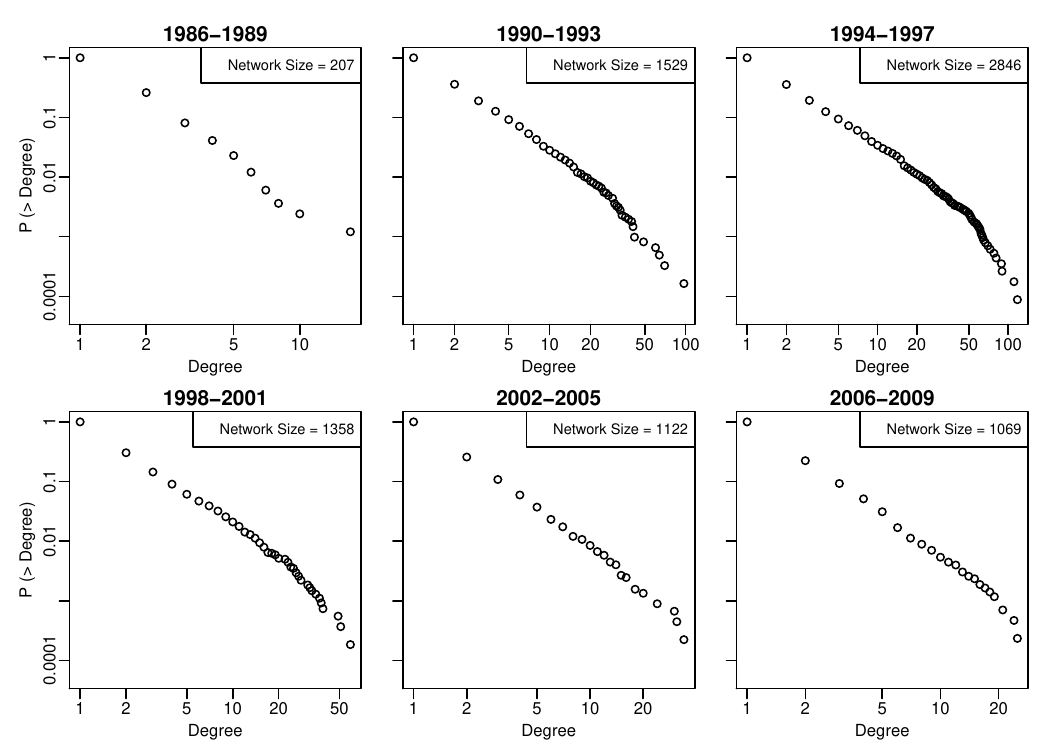}
  \end{center}
  \vspace{-4pt}
  \caption[]{\textbf{Degree distribution in the pooled R\&D network.}
    Complementary cumulative degree distributions of the pooled R\&D
    network in six sub-periods. \textit{Note:} the insets in the top
    right corner show the average network size in each of the
    sub-periods.}
  \label{fig:degree_distr}
\end{figure}

Table~\ref{table:ddistr} shows the first four moments of the degree
distribution of the pooled network in each sub-period. In all periods examined,
the degree distribution displays high variance associated with high
right-skewness and excess kurtosis. In addition, the p-values of the
Kolmogorov-Smirnov test show that the degree distributions of the
pooled network are extremely far from the Normal benchmark.  In
particular, the very high values of the kurtosis coefficient
(especially in the period 1994-1997) are indicative of heavy tails in
the degree distribution, which in turn imply the
presence of network ``hubs'' concentrating a high number of
alliances. This is also confirmed by the visual analysis of such
distributions, reported in Figure \ref{fig:degree_distr}.

\begin{table}[h!]
  \footnotesize
 
  \centering
  \begin{tabular*}{0.90\linewidth}{@{\extracolsep{\fill}}l*{6}{@{ }r}}
    \hline
    \textbf{} & \textbf{1986-1989} & \textbf{1990-1993} & \textbf{1994-1997} & \textbf{1998-2001} & \textbf{2002-2005} & \textbf{2006-2009} \cr
    \hline
    Mean &  1.44 &   2.25 &   2.40 &  1.91 &  1.57 &  1.47 \cr 
    SD &  1.09 &   3.77 &   4.75 &  2.87 &  1.70 &  1.42 \cr 
    Skewness &  5.85 &   8.93 &   9.65 &  8.04 &  7.74 &  7.03 \cr 
    Kurtosis & 62.06 & 134.90 & 141.27 & 99.42 & 98.20 & 77.81 \\
    [0.6ex] 
    KS test $p$-value & $<10^{-15}$ & $<10^{-15}$ & $<10^{-15}$ & $<10^{-15}$ & $<10^{-15}$ & $<10^{-15}$ \\
    \hline
  \end{tabular*}
  \caption{Degree distribution statistics and $p$-values of
    Kolmogorov-Smirnov (KS) test for the pooled R\&D
    network.}
  \label{table:ddistr}
\end{table}

Furthermore, Table \ref{table:ddistr} shows that all the
four moments of the degree distribution increase in the first years of
the sample, reaching a peak in the 1994-1997 period, and then
decrease again. This indicates that the ``golden age'' of R\&D
networks has been
characterized by more alliance activity per firm, but also by more
alliance inequality.

The degree distributions of the sectoral R\&D networks display
patterns that are similar to those of the pooled R\&D
network.\footnote{These results are not shown here, but are available
  from the authors upon request.} In particular, all sectoral degree
distributions are characterized by high variance associated with
significant skewness and kurtosis in all sub-periods. Again, also in sectoral
networks, firms have on average more collaborators during the ``golden
age'' of alliance activity (1994-1997) but also more unequal alliance
activity.

The previous analysis thus suggests the significant presence of heavy tails in both the
pooled and sectoral degree distributions. To get an estimate of the
``heaviness'' of those tails from a non-parametric point of view, we
compute the Hill Estimator \citep{hill1975} (HE), a tool commonly used
to study the tails of economic data (see the Appendix for more details).
It is important to recall here that the theoretical HE value predicted by the
preferential attachment model of \citet{barabasi99:_emerg} is 3. A
value of the HE lower than 2 indicates an extremely heavy-tailed
distribution -- ``super heavy-tailedness''. At the other extreme, a
value higher than 4 is indicative of degree distributions whose
fat-tail property is not very pronounced -- ``sub heavy-tailedness''.

\begin{table}[h!]
  \scriptsize
  
  \centering
  \begin{tabular*}{0.99\linewidth}{@{\extracolsep{\fill}}l*{6}{@{ }r}}
    \hline
    & \textbf{1986-1989} & \textbf{1990-1993} & \textbf{1994-1997} & \textbf{1998-2001} & \textbf{2002-2005} & \textbf{2006-2009} \cr
    \hline

    \textbf{Pooled} & 3.43 & 2.36 & 2.30 & 2.52 & 2.86 & 2.92 \cr 
    Pharmaceuticals & 4.72 & 3.10 & 2.29 & 2.53 & 2.95 & 3.02 \cr 
    Computer Software & 3.24 & 2.16 & 2.07 & 2.98 & 2.51 & 3.29 \cr 
    Electronic Components & 3.33 & 2.81 & 2.11 & 2.16 & 2.42 & 3.03 \cr 
    Computer Hardware & 2.82 & 2.09 & 1.96 & 2.60 & 2.33 & 4.28 \cr 
    Medical Supplies & - & 3.10 & 2.63 & 3.68 & 4.20 & 4.18 \cr 
    Communications Equipment & 5.10 & 3.00 & 1.97 & 1.95 & 2.30 & 3.13 \cr 
    Laboratory Apparatus & 4.09 & 2.19 & 3.39 & 2.57 & 2.68 & 4.26 \cr 
    Motor Vehicles & 5.07 & 3.65 & 2.06 & 3.10 & 2.85 & 2.73 \cr 
    Inorganic Chemicals & 3.29 & 2.59 & 2.22 & 3.28 & 3.08 & 3.45 \cr 
    Aircrafts and parts & 3.15 & 2.77 & 3.30 & 4.19 & 2.79 & 6.47 \cr 

    \hline \hline
  \end{tabular*}
  \caption{Hill Estimators (HE) for degree distributions in the pooled
    and the sectoral R\&D networks. \textit{Note}: missing values refer to sectors with not
    enough observations. For each degree distribution, we also compute the p-value of a Kolmogorov-Smirnov test; our null hypothesis is that the data are drawn from a distribution having the fitted HE value. We find that, in \textit{all} cases, the null hypothesis cannot be rejected, thus supporting the significance of all reported HE values (see Appendix for more details).}
  \label{table:he}
\end{table}

Table \ref{table:he} reports the values of the Hill estimator (HE) for
both the pooled and the sectoral R\&D networks in all the considered
time periods. Starting with the pooled network, we observe that the HE
first decreases, reaching a minimum in the golden-age period
1994-1997, and then increases again. The values of the HE computed on
the sectoral R\&D networks reveal a rise-and-fall pattern similar to
the one detected in the pooled network (see Table
\ref{table:he}). Again, most sectors display fatter tails in the
periods of higher alliance activity.  All this shows that the degree
of tail-heaviness undergoes a rise-and-fall dynamics similar to the
other network measures discussed so far. Moreover, Table
\ref{table:he} also shows
that, in all periods, the HE mostly ranges between 2 and 4. This
rules out both super and sub heavy-tailedness. Finally, in all
time periods, except the first and the last one, the values of the HE
are significantly below 3, and the minimum is reached in the golden
age period 1994-1997 (2.30, for the pooled R\&D network).
\footnote{As an additional check, we perform a series of Kolmogorov-Smirnov (KS)
tests. The corresponding p-values are reported in the Appendix. For
all networks in all time periods, the KS test could not reject the
hypothesis that the original data are drawn from a distribution having
the fitted HE value.} In line with previous empirical works
\citep[e.g.][]{powell2005network}, this finding indicates that in
those periods the degree distributions of R\&D networks are
not consistent with the preferential-attachment model -- being their
tails ``fatter'' than what predicted by that model. Accordingly, a
simple accumulative advantage process is not enough to fully account
for the observed heterogeneity in R\&D networks.

\medskip

Overall, the above results show that the process of R\&D alliance formation has been
characterized by huge and persistent cross-firms heterogeneity in
terms of number of alliances. We now turn to investigate how partners'
choices of firms having similar characteristics are correlated. One example of this is
provided by the visualization of sectoral networks, presented in Figure
\ref{fig:sec_net}. Indeed, the figure shows that firms in the
pharmaceutical sector have displayed a stronger preference towards
alliances with firms in the same sector, whereas this has
been less the case in the other studied sectors. Such a preference
for the formation of alliances with actors of similar type (e.g.~of
the same sector) is an instance of \textit{structural homophily}. The
existing literature on R\&D networks has explained how homophily may
reflect a series of technological as well as organizational drivers. It
may for instance be driven by the similarity in knowledge bases, and
therefore by the need to establish connections with firms with whom it
is possible to ``communicate'' and therefore absorb knowledge
\citep[see
e.g.][]{powell2005network,gulati2012:_rise_and_fall_of_small_world,cowan2009knowledge}. Homophily
may also be generated by the preference for forming relationships with
firms having similar organizational structure and thus displaying the same
capacity in managing alliances
\citep[e.g.][]{rosenkopf2008investigating}. Finally, and especially
for ICT-related sectors, it can be due to the need to establish 
technological standards
\citep[][]{rosenkopf2008investigating,gulati2012:_rise_and_fall_of_small_world}.

One indicator of structural homophily traditionally used in the
network literature is the degree of
network assortativity, measured by the correlation in degree across partners
\citep[\textit{assortativity mixing
  coefficient}; see][]{pastor-satorras01:_dynam_correl_proper_inter,newman02:_assor_mixin_in_networ}.
A network is assortative if it is characterized by a positive correlation
across the degrees of linked nodes. Assortative networks display high
homophily as firms tend to be connected to firms with similar
degree. At the other extreme, disassortative networks have negative
degree-degree correlation, i.e. nodes tend to be connected to nodes with
dissimilar degree. \citet{newman03:_mixin_patter_in_networ} finds that
technological networks, such as the Internet, are disassortative while
social networks, such as the network of scientific co-authorships, are
assortative.

We compute the assortativity mixing coefficient $r$, defined in
\citet{newman03:_mixin_patter_in_networ}, on both the pooled and the
sectoral R\&D sub-networks (see the Appendix for more details). Similar to the previous section, the whole
observation period is divided into six sub-periods of 4 years
each and all the observations of every firm's degree are taken
together within each sub-period. The degree correlation coefficients
are then computed for each sub-period. The results are reported in
Table~\ref{table:assortativity}.

\begin{table}[h!]
  \scriptsize
  \centering
  \begin{tabular*}{0.99\linewidth}{@{\extracolsep{\fill}}l*{6}{@{ }r}}
    \hline
    & \textbf{1986-1989} & \textbf{1990-1993} & \textbf{1994-1997} & \textbf{1998-2001} & \textbf{2002-2005} & \textbf{2006-2009} \cr
    \hline
 
    \textbf{Pooled} &  0.018 &  0.128 &  0.115 & 0.210 &  0.262 &  0.004 \cr 
    Pharmaceuticals &  0.271 &  0.328 &  0.336 & 0.306 & -0.046 & -0.051 \cr 
    Computer Software & -0.113 &  0.010 &  0.009 & 0.104 &  0.505 & -0.027 \cr 
    Electronic Components &  0.209 &  0.112 &  0.068 & 0.201 &  0.316 &  0.595 \cr 
    Computer Hardware & -0.106 & -0.020 & -0.047 & 0.033 &  0.186 &  0.317 \cr 
    Medical Supplies & -0.100 &  0.545 &  0.476 & 0.259 &  0.045 & -0.004 \cr 
    Communications Equipment &  0.188 &  0.057 &  0.036 & 0.312 &  0.407 &  0.303 \cr 
    Laboratory Apparatus & -0.046 &  0.312 &  0.207 & 0.396 &  0.487 &  0.059 \cr 
    Motor Vehicles & -0.151 &  0.337 &  0.317 & 0.383 &  0.386 &  0.864 \cr 
    Inorganic Chemicals &  0.245 & -0.052 &  0.233 & 0.297 &  0.422 & -0.153 \cr 
    Aircrafts and parts &  0.173 &  0.195 &  0.417 & 0.291 &  0.423 &  0.556 \cr 

    \hline \hline
  \end{tabular*}
  \caption{Assortativity mixing coefficient in the pooled and the
    sectoral R\&D networks (SIC codes are in
    brackets).}
  \label{table:assortativity}
\end{table}

\begin{figure}[t]
  \begin{center}
    \includegraphics[width=1\textwidth,angle=0]{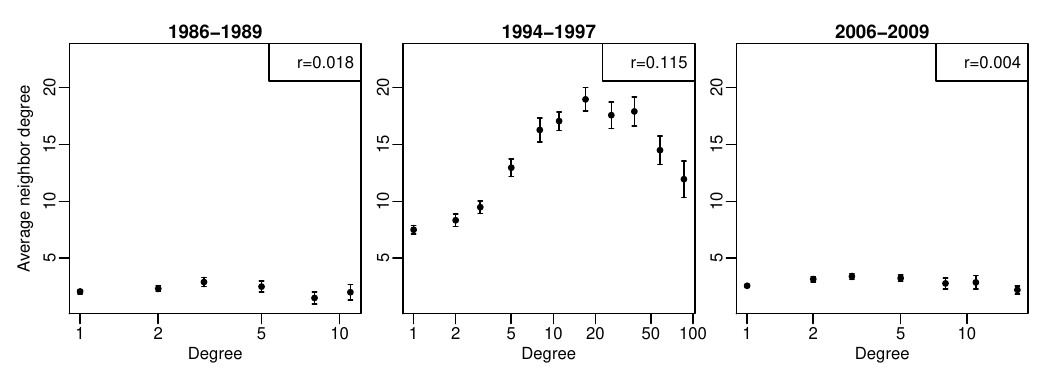}
  \end{center}
  \vspace{-10pt}
  \caption[]{Average neighbors' degree as a function of firm's degree
    in the pooled R\&D network. The whiskers measure +/-2*standard
    deviation from the mean. \textit{Note}: on the top-right corner of
    each plot we report the corresponding value of the assortativity
    mixing coefficient in the sub-period under analysis.}
  \label{fig:assortativity}
\end{figure}

We find that both the pooled and the sectoral R\&D networks are
generally assortative. Correlation coefficients are low but positive
during the whole observation period (see Table
\ref{table:assortativity}) and in line with similar works in the
complex networks literature
\citep[e.g.][]{newman03:_mixin_patter_in_networ}. This means that, on
average, high-centrality (low-centrality) firms tend to connect to
other high-centrality (low-centrality) firms.  Moreover, although
correlation coefficients tend in general to be higher in the golden
age phase, no particular rise-and-fall dynamics seems to be
present.\footnote{We also calculated the assortativity network
  coefficient on the network including service sectors (e.g. business
  services, management and consulting) as well as universities. Such a
  network is still assortative when alliances are considered
  irrespective of the sector of the partners, but it turns out to be
  disassortative when sectoral networks are considered. Such
  differences hint to a peculiar role played by universities and
  service firms in the process of formation of R\&D alliances, a topic
  that we leave to future research.}

To shed more light on the mechanics of alliance behavior generating
assortativity, we plot in Figure \ref{fig:assortativity} average
neighbors' degree as a function of firm's degree, for three different
periods of our sample. In such a plot, assortativity should correspond to
an increasing relation between the two variables. In contrast, the plots
show quite neatly that such an increasing relation is present, at
best, only in the golden age period (1994-1997). No relation is instead present at the beginning and the end of the observation
period (respectively, 1986-1989 and 2006-2009). Moreover, in the golden
age, the relation between average neighbors' degree and firm's
degree is highly non-linear. More specifically, hubs (i.e. nodes
with degree larger than 20) tend on average to connect
to nodes with intermediate degrees rather than to other hubs. It
follows that the assortativity one observes in the pooled network masks quite different
alliance behaviors from different types of firms. On the one hand, the behavior of
firms having a low or intermediate degree tends to be dictated by
homophily considerations, as indicated by the search for partners
having similar degree. On the other hand, hubs exhibit a different strategy,
mainly forming partnerships with more
peripheral firms -- in terms of number of alliances.

\section{From small worlds to core-periphery architectures}
\label{sec:small-worlds-comm}

One basic fact about R\&D network dynamics that we have spotted in
Section \ref{sec:basic-netw-prop} is the emergence of dense giant
components in the peak years of alliance formation. Large network
components allow high connectedness and thus increase the number
of knowledge sources that single firms can reach, either directly or indirectly. In
this section we turn to analyze more in depth how these components
were organized. This is important because different component
architecures reflect different models of alliance
formation. Accordingly, their analysis sheds further light on which
type of processes governs the evolution of R\&D networks in our
sample.
In this respect, \textit{small worlds} are one type of network architecture that has received
significant attention both in the empirical and theoretical literature
on R\&D networks  \citep[see
e.g.][]{uzzi2007small, fleming07:_small_world_and_region_innov,
  gulati2012:_rise_and_fall_of_small_world,cowan2009knowledge,cowan2004network}. A
network is a small world if it is characterized by two key features
\citep{strogatz98:_collective_dynamics_small_world_networks}:
\textit{a)} high local clustering, i.e. a structure where the
neighbors of a node are in their turn connected among themselves, and
\textit{b)} low average path length, i.e. the existence of short paths connecting a node to any other node in the network.\footnote{Local
  clustering is defined as the number of existing
  links between the neighbors of a focal node, divided by the number
  of all possible links between these neighbors. Average path length
  is defined as the average of all shortest distances, i.e. the lowest
  number of links that must be traversed to connect every pair of
  nodes in the network. See Appendix for further
  details.}

Small worlds are typical of many
technological and social domains \citep[see][]{newman_book}. In the
context of R\&D collaborations, they may emerge as a result of a
tension between homophily and diversity in the search of partners.
On the one hand, densely connected components can be generated by the need to
ensure trust among partners and discourage non-cooperative
behavior \citep{gulati1995familiarity}. Likewise, they can occur
because of technological
similarity, when firms are trying to exploit scale economies in their
search for innovation
\citep{gulati2012:_rise_and_fall_of_small_world}. On the other hand,
low average path length can be the result of the effort of some firms
to establish bridging ties across different communities,
in order to get access to new ideas and sources of knowledge, and
thus dampen the possible adverse effects on innovation of the  
redundancy characterizing knowledge exchanges in
closely interconnected clusters
\citep{sytch08:_where_do_broker_come_from,rosenkopf2003overcoming}. 

We compute the small world
coefficient, defined as the ratio between the clustering coefficient
and the average path length of the network \citep[][see also Appendix
for more
details]{strogatz98:_collective_dynamics_small_world_networks}, on
the pooled and the sectoral R\&D networks, and compare it to the small
world coefficient that would emerge in randomly generated networks
having the same size as the empirical ones.  The results of our computations are listed in
Table~\ref{table:sw}. Once again, the results are presented for six
different sub-periods.\footnote{The small world quotient is computed
  separately for every year during the whole observation period, in
  both the pooled and the sectoral R\&D networks, and then averaged
  within each sub-period. We do not aggregate the observations inside
  every time period, because the small world quotient is a global
  network measure, and not an ego-network measure centered around
  single nodes.}
  Values higher (lower) than one indicate
that the degree of small-worldliness of the empirical network under
scrutiny is higher (lower) than what would be predicted by a random network
\begin{table}[h]
  \scriptsize
  \centering
  \begin{tabular*}{0.99\linewidth}{@{\extracolsep{\fill}}l*{6}{@{ }r}}
    \hline
    & \textbf{1986-1989} & \textbf{1990-1993} & \textbf{1994-1997} & \textbf{1998-2001} & \textbf{2002-2005} & \textbf{2006-2009} \cr
    \hline

    \textbf{Pooled} & 1.38 & 42.93 & 90.43 & 32.08 & 9.81 & 2.55 \cr 
    Pharmaceuticals & 0.00 & 19.83 & 44.87 & 20.75 & 4.06 & 2.36 \cr 
    Computer Software & 0.90 & 17.56 & 39.34 & 12.08 & 4.86 & 0.35 \cr 
    Electronic Components & 1.47 & 12.13 & 20.53 & 11.72 & 6.79 & 2.06 \cr 
    Computer Hardware & 0.42 & 14.52 & 25.56 & 10.22 & 4.50 & 0.00 \cr 
    Medical Supplies & 0.00 &  2.82 &  7.03 &  1.46 & 0.39 & 0.23 \cr 
    Communications Equipment & 1.78 &  8.28 & 15.34 &  8.10 & 4.19 & 1.46 \cr 
    Laboratory Apparatus & 0.00 &  5.25 &  5.58 &  2.95 & 1.84 & 0.64 \cr 
    Motor Vehicles & 0.99 &  4.17 &  7.15 &  4.28 & 3.09 & 1.18 \cr 
    Inorganic Chemicals & 1.29 &  3.66 &  6.38 &  1.24 & 1.23 & 0.00 \cr 
    Aircrafts and parts & 0.83 &  4.50 &  5.38 &  3.02 & 1.93 & 2.09 \cr

    \hline \hline
  \end{tabular*}
  \caption{Small world quotient of pooled and sectoral R\&D
    networks, for the \textit{giant
      component}. The values are
    averages within each sub-period.}
  \label{table:sw}
\end{table}

First, Table \ref{table:sw} shows that small worlds are a universal
characteristic of both the pooled and the sectoral networks that we
analyze. The ratios in the table are in general higher than one, especially
in  the periods of more intense alliance activity (from 1990 to 2001). Second, periods of
more intense alliance activity are also characterized by significant
cross-sectoral heterogeneity, in terms of small-worldliness.
Small world ratios are indeed considerably higher in
sectors like Pharmaceuticals and Electronic
Components than in sectors like Laboratory Apparatus, and Aircrafts
and Parts (cf. Table \ref{table:sw}). Third, the evolution of the small
world quotient exhibits a universal marked rise-and-fall pattern over
time. Both at the pooled and sectoral level the small world property is basically absent at the beginning of
our observation period (1986-1989), then it increases, reaching a
peak in the ``golden age'' (1994-1997), and then disappears
again. 

The above findings generalize previous results in the empirical
literature that were limited to single industries or geographical areas
\citep[e.g.][]{gulati2012:_rise_and_fall_of_small_world,fleming07:_small_world_and_region_innov,fleming07:_evolut_of_inven_networ_in},
by showing that, indeed, small worlds are a robust property of the
global network of R\&D alliances as well as of many sectoral
networks. In addition, they show that small worlds have displayed
a rise-and-fall dynamics similar to other measures discussed so far. As
it is argued at more length in
\citet{gulati2012:_rise_and_fall_of_small_world}, this evolution can be
explained by the fact that the same forces leading to small worlds
(the above mentioned tension between homophily and 
diversity in the search for partners) have also set the premises for their own
destruction. On the one hand, technological similarity and
self-reinforcing trust may lead actors within clusters to refuse the
entry of new organizations. In addition, increasingly homogeneous
clusters lacking diversity may also become less attractive to
newcomers searching for new ideas. On the other hand, the diversity
underlying the establishment of bridging ties may disappear as
the small world matures, because knowledge exchanges render the knowledge bases of firms more
homogeneous over time, even in different
clusters \citep{cowan2004network}. Accordingly, the incentive to keep
bridging ties may decay and
the network may become fragmented. Of course this fragmentation may lead to knowledge heterogeneity across clusters, which eventually   will pave the way for a new cycle to emerge. A similar rise-and-fall dynamics can
also be explained by models based on the idea of \textit{multiconnectivity}
\citep{powell2005network}, according to which the rate of knowledge growth in a network is
determined by the sources of knowledge to which a firm has either
direct or indirect access (in other words, the degree of network
connectedness). In such a framework, when the network is 
sparse, creating highly interconnected clusters and bridging ties is a
way to increase connectivity within the system, and therefore to create
multiple knowledge paths. However, once the network is dense and
connected, more paths can be created by reinforcing ties within
existing clusters, rather than keeping connections with more distant
partners. The result is, again, a stronger incentive to remove
bridging ties, resulting in the fragmentation of the network and the disappearance
of the small world structure \citep[see][for an example of model based
on the idea of multiconnectivity and generating similar dynamics]{KoBaNaSchJEBO}.

The above analysis provides important insights into how R\&D
alliances are organized within connected networks. At the same time,
many diverse network architectures may co-exist under the umbrella of
``small worlds'', which -- by themselves -- do not place strong restrictions on
the class of possible generating alliance mechanisms. In
particular, the small world property can be displayed both by a network
where many clusters, populated by firms with relatively homogeneous degree, are
sparsely connected among themselves, and by a ``core-periphery'' network, where
a core of densely inter-connected firms is linked to a periphery of
firms having only a few ties. Interestingly, our analysis performed in Section
\ref{sec:homoph-hetero-alli-behav} has shown that the degree
distribution of the R\&D networks is fat-tailed, a typical property
of core-periphery structures.

A generalization of the concept of core-periphery architectures is represented by 
the so-called \textit{nestedness}. Developed and studied for the first time
in the domain of ecological networks \citep[see][]{bascompte2003nestedness},
nestedness quantifies the presence of hierarchies in a network's topology.
In a nested network, the set of partners of a node (its neighborhood) is contained in the set
of partners of nodes with higher degree. In that respect, nested
networks are a more general notion than standard single-core
single-periphery structures, as they may feature several cores
connected to several peripheries. A number of works in the domain
of inter-firm networks have pointed out that nested networks can facilitate knowledge growth in
models based on multiconnectivity
\citep[e.g.][]{konig2012efficiency}. In addition, nested structures can arise in
R\&D network growth models where alliance formation is determined by
competition in the search for centrality \citep[e.g.][]{koenig2013nestedness}.

Nestedness can be quantified through an indicator, that we call the \textit{nestedness coefficient} of the network.
In Table \ref{table:nestedness} we report the values of
such coefficient for the pooled and the
sectoral R\&D networks, across different periods.\footnote{We have also
  computed the values of a different network indicator, namely the core-periphery
  coefficient $C_{cp}$ suggested by \citet{holme2005core}. Our results
  are robust to such a different choice of indicator.}
The coefficients in the table are normalized so that 1
corresponds to a fully nested network, whereas 0 corresponds to
a completely random network (see the Appendix for more details on the computing procedure).

\begin{table}[h!]
  \scriptsize
  \centering
  \begin{tabular*}{0.99\linewidth}{@{\extracolsep{\fill}}l*{6}{@{ }r}}
    \hline
    & \textbf{1986-1989} & \textbf{1990-1993} & \textbf{1994-1997} & \textbf{1998-2001} & \textbf{2002-2005} & \textbf{2006-2009} \cr
    \hline

    \textbf{Pooled} & 0.791 & 0.996 & 0.999 & 0.996 & 0.988 & 0.994 \cr 
    Pharmaceuticals 		& 0.706 & 0.983 & 0.993 & 0.990 & 0.988 & 0.994 \cr 
    Computer Software	 	& 0.823 & 0.991 & 0.997 & 0.979 & 0.930 & 0.805 \cr 
    Electronic Components	 	& 0.778 & 0.979 & 0.991 & 0.980 & 0.950 & 0.820 \cr 
    Computer Hardware	 	& 0.804 & 0.991 & 0.995 & 0.982 & 0.921 & 0.662 \cr 
    Medical Supplies 		& 0.680 & 0.838 & 0.926 & 0.701 & 0.750 & 0.775 \cr 
    Communications Equipment	& 0.464 & 0.955 & 0.988 & 0.966 & 0.893 & 0.585 \cr 
    Laboratory Apparatus	 	& 0.728 & 0.879 & 0.945 & 0.911 & 0.799 & 0.732 \cr 
    Motor Vehicles	 		& 0.630 & 0.835 & 0.949 & 0.898 & 0.813 & 0.651 \cr 
    Inorganic Chemicals		& 0.561 & 0.925 & 0.924 & 0.762 & 0.678 & 0.805 \cr 
    Aircrafts and parts	 	& 0.688 & 0.848 & 0.874 & 0.827 & 0.801 & 0.570 \cr 

    \hline \hline
  \end{tabular*}
  \caption{Nestedness coefficients for the pooled and the sectoral
    R\&D networks. The values
    are averaged in six sub-periods.}
  \label{table:nestedness}
\end{table}

Table \ref{table:nestedness} shows that the normalized nested coefficients
are always very high, both for the pooled and sectoral networks. In
addition, in the golden age (1994-1997), all values
are extremely close to one, indicating the presence of fully nested
structures.\footnote{The only exception is represented by the
  coefficient of the sector ``Aircrafts and Parts'', whose value
  is nonetheless very high (0.874). Moreover, we have found that most nestedness coefficients
  found in our R\&D networks (and \textit{all} coefficients in the ``golden age'') are significantly different from the
  average values of a set of random networks used as benchmark; see Appendix
  for more details.} Finally, similar to other
network measures discussed in this paper, also the 
nestedness follows a rise-and-fall pattern over time, in both the pooled and the sectoral networks. In the next
section, we focus on the implications arising from the
descriptive analysis performed so far, and we investigate
the validity of different statistical models on R\&D alliance formation
via regression analyses.

\section{Investigating the determinants of alliance formation}
\label{sec:alliance_formation}

In the previous sections, we have shown that the dynamics of R\&D networks in
the years 1986-2009 is characterized by two distinct growth
phases. During the first of them (the ``rise'' phase, from 1986 to
1997), the number of R\&D alliances boosted and gave rise to highly connected
network components displaying significant unevenness and moderate
homophily in terms of alliance activity (as indicated, respectively,
by fat-tailed degree distributions and low but positive assortativity
mixing coefficients, cf. Section
\ref{sec:homoph-hetero-alli-behav}). In addition, connected components
were organized into core-periphery architectures displaying the small
world property (see Section \ref{sec:small-worlds-comm}). In the
second phase (the ``fall'' phase, from 1998 to 2009) the rate of
alliance activity declined, and the R\&D network became more
fragmented into smaller components not displaying the properties of
the previous phase. Finally, the above described network evolution has
been \textit{universal}, the same growth patterns emerging both when
alliances were considered regardless of the firms' sectors and
when different sectoral networks were analyzed.
 
We now turn to a statistical analysis of the determinants of R\&D
alliance formation. Following the previous empirical literature on such alliance
determinants \citep[e.g.][]{powell2005network,rosenkopf2008investigating},
we assume that the process of R\&D alliance formation is highly
path-dependent and that the topological characteristics of the
existing network structures and the rules of attachment among its
constituents determine the choice of future partners, thus shaping the
evolution of the network itself. 

Network characteristics can be firm-specific (e.g. the degree of a
potential partner in the network) or relate to the structure of the
component to which potential partners belong to (e.g. the number of
paths within the component). Moreover, they capture important technological as well as organizational
and social factors driving alliances. For instance, the position of a firm in the network
(e.g. its centrality) captures its access to multiple knowledge sources,
or a better experience in managing R\&D collaborations (see Section
\ref{sec:homoph-hetero-alli-behav}). In addition, the fact of being part of a highly
interconnected cluster can improve trust and communication among
partners (cf. Section \ref{sec:small-worlds-comm}).

As far as rules of attachment are concerned, we focus
on the following hypotheses.

\bigskip
\noindent \textit{Hypothesis 1: Accumulative Advantage}. The probability of forming
an alliance with a firm increases with the centrality of that firm in
the network. 

\bigskip

\noindent\textit{Hypothesis 2: Structural Homophily (or Diversity)}. The
probability of an alliance between two firms increases with their 
similarity (diversity). 

\bigskip
\noindent \textit{Hypothesis 3: Multiconnectivity}. The
probability of forming an alliance with a firm increases if that
firm allows to reach other firms in the network through multiple
independent paths. 

\bigskip
Similarly to topological characteristics, rules of attachment are
stylized representations of different evolutionary drivers underlying
the formation of alliances. For instance, accumulative advantage captures
the presence of increasing returns in the alliance process, i.e. a
situation where
firms that are already more visible in the network are also able to
attract more partners. Likewise, processes based on structural
homophily reflect settings where firms search for similar partners in
terms of technological (e.g. same sector), spatial
(e.g. same geographical area) or network (e.g. being
part of the same alliance cluster) characteristics. With these partners,
communication and trust occur faster, facilitating 
the exchange and the absoption of knowledge and the sharing of resources \citep[see
e.g.][]{gulati07:_manag_networ_resour}. In contrast, structural
diversity mainly reflects firms' exploratory search for novel and
different knowledge paths \citep[see
e.g.][]{rosenkopf2003overcoming,rosenkopf2008investigating,rowley00:_redun_gover_struc}.
Finally, multiconnectivity reflects alliance behavior in contexts
where innovation is driven by knowledge recombination, and thus where it
becomes of fundamental importance to have access to multiple knowledge sources
\citep[see][]{powell2005network}.

We have selected the above mentioned attachment rules partly because of the attention
that they have received in previous empirical studies \citep[see
e.g.][]{powell2005network,rosenkopf2008investigating,gulati1995social}
and partly because of the hints stemming from our previous descriptive
analysis. Indeed, fat-tailed degree
distributions and core-periphery structures (cf. Sections
\ref{sec:homoph-hetero-alli-behav} and \ref{sec:small-worlds-comm})
can be the result of an accumulative advantage process where firms with a
more central position in the network (e.g. with higher degree) are
able to attract more partners.\footnote{However, our results also
  indicate that such a process is different from a standard
  preferential attachment dynamics \'{a} la Barabasi and Albert (see
  Section \ref{sec:homoph-hetero-alli-behav}).} Likewise, small worlds (and their
rise and fall), can be the result of processes based on structural
homophily or multiconnectivity. Finally, multiconnectivity can also
account for the high network connectedness documented in Section
\ref{sec:basic-netw-prop}.

In what follows we perform a statistical analysis of the above
described attachment rules. As the characteristics of network dynamics
appear to be universal across sectors and scale of aggregation, we
focus only on the pooled network.  Our observation unit is not a firm,
but a \textit{dyad of firms} -- i.e. a pair of potential partners in
the R\&D network.  Moreover, given the importance of firm entry in the
network dynamics, we perform regressions on two different sets of firm
dyads (also called ``risk sets''): one where both potential partners
are incumbents, i.e. they have at least one alliance in their history,
and one where at least one firm is an entrant firm, i.e. it has no
previous alliance activity \citep[see also][for a similar
exercise]{rosenkopf2008investigating}. Finally, we analyze whether
different attachment rules are at work in the phase of rise and in the
one of fall. For this reason, we run separate regressions for the period
1986-1997 and the period 1998-2009. We begin by describing the
variables used in our regressions. Next, we present the employed
statistical methodology and discuss the results of our analysis.

\subsection{Variables}
\label{sec:vari-stat-meth}    

\paragraph{Dependent variable.}
We record the alliance history of all firm dyads in the network, in
each year, from 1986 to 2009. Given the huge number of potential
observations, we exclude from the analysis firms that have been
involved in less than 5 alliance events during the entire observation
period. Even after this censoring, we still obtain a sample with over one million dyad-year
observations.  Next, for each dyad-year observation, we record a binary
dependent variable, \textit{Alliance formation}, expressing whether
the considered dyad forms an R\&D alliance in the considered year.
Consistently with our network representation (cf. Section
\ref{sec:data}), alliance consortia are coded as multiple two-party
alliances between each pair of members of the consortium, and
reverse-ordered dyads are excluded from the sample 
(our R\&D network is undirected).

As explained before, we select two different ``risk sets'' on which we perform our
dyadic regressions, the distinction being based on the alliance
history of the potential partners: A) incumbent
dyads, where both firms have already been involved in at least one
alliance;  B) mixed dyads, where one firm is incumbent and the other
one is a new-entrant in the R\&D network.  Moreover, we divide our
sample in two observation periods, rise (1986-1997) and fall
(1998-2009).  Accordingly, we run four different batteries of
regressions: 1) on incumbent dyads in the rise phase; 2) on incumbent
dyads in the fall phase; 3) on mixed dyads in the rise phase; 4) on
mixed dyads in the fall phase.

\paragraph{Independent variables.}
All of our dyadic independent variables are computed in the year
preceding the dyad-year observation under examination. Some of them
relate to structural features of the involved firms, whilst others
relate to their network
characteristics.
To compute the latter, we
construct year by year networks using the same procedure described in Section
\ref{sec:data}.
Following \citet{powell2005network} and
\citet{rosenkopf2008investigating}, we focus on structural features
and network characteristics identifying the different attachment
rules that are the object of our investigation. 

\medskip 

\textit{Accumulative advantage.} We assume that the formation of
a new alliance will most likely involve the most central firms in
the network. This is captured by the variable \textit{Joint
  centrality} for incumbent dyads, which expresses the average of the degree 
centrality of the two firms, and by the variable \textit{Incumbent
  centrality} for mixed dyads, which expresses the degree centrality of
the incumbent firm in the dyad.\footnote{Given the skewed nature of the degree centrality distributions (see Section \ref{sec:homoph-hetero-alli-behav}), and the multiplicative -- rather than additive -- mechanism behind them, we take the logarithm of the variables \textit{Joint centrality} and \textit{Incumbent centrality}. We have also
tested models with other centrality measures -- such as 
closeness, betweenness and eigenvector
centrality. Results were robust to the use of these alternative
measures. Notice that all these measures are strongly
correlated with each other and give rise to collinearity if
used together in the same model.} We expect the probability of
an alliance between two firms to be positively
correlated with both joint and incumbent centrality if the alliance
formation process is driven by accumulative advantage. 

\medskip 

\textit{Structural homophily and diversity.} This group of variables
quantifies the extent to which the two firms in the dyad are similar,
with respect to both network related and structural, non-network related, 
characteristics.
The dummy variables \textit{Same nation} and \textit{Same SIC}
measure, respectively, whether the firms are registered in the same
country and whether they have the same SIC code (at a 3-digit level),
as recorded in the SDC dataset.  They capture
geographical and technological similarity. The latter is also captured
by the variable \textit{Technological
  distance}, a real number that expresses how different the technological
positions of the two companies are, computed through their
patents. For that, we use data
provided by the NBER patent database  \citep[see][]{hall01:_nber_paten_citat_data_file}, listing
all patent applications in the US and their respective categories, by US and non-US firms, from 1976
to 2006. See the Appendix for more details on the calculation of this measure.
The last three variables are computed in the same way for
both the incumbent and mixed dyads risk set. If alliance formation is
based on homophily (resp. diversity) we expect the probability
of an alliance to be positively (resp. negatively) correlated with sector and nation
dummies, and negatively (resp. positively) correlated with
technological distance.  Network similarity is defined only for
incumbent dyads (mixed dyads include firms that are not
part of the network yet, and for which network measures are by definition 
null), and it is measured by the
variables \textit{Centrality inequality}, \textit{Inverse path length}, and \textit{Common
  neighbors}. The first variable expresses
the ratio between the degree centralities of the two firms (the
largest divided by the smallest). The second variable expresses the network distance (as opposed
to technological distance) of the two firms. Following the
approach adopted in Section \ref{sec:small-worlds-comm}, we define the
shortest path length as the number of links in the network that have
to be traversed in order to connect the two firms. This is an integer
number ranging from 1 (if the firms are directly connected) to
infinite (if the firms belong to disconnected network
components); accordingly, the value of \textit{Inverse path length}
ranges from 0 (disconnected firms) to 1 (directly connected
firms).  \textit{Common neighbors} expresses the
number of alliance partners that the two firms have in common. 
Positive (negative) coefficients of these two variables indicate the
presence of structural homophily (diversity) in the firms' attachment rules. 

\medskip 

\textit{Multiconnectivity.} As shown by \citet{powell2005network},
partners that allow a firm to reach many other firms through multiple
independent paths are the most attractive alliance partners. For
incumbent dyads, we capture this idea via the use of the variable
\textit{Average multi-path growth}, that measures the average largest
eigenvalue of the connected components to which the two potential
partners belong.  As explained in \citet{KoBaNaSchJEBO}, the largest
eigenvalue of the adjacency matrix of a connected component in a
network is directly proportional to the number of multiple independent
paths within the component.\footnote{\citet{powell2005network} use a
  different measure of multiconnectivity, namely the k-coreness. We
  believe that the largest eigenvalue provides a better description of
  the idea of multiconnectivity, as it directly measures the growth of
  multiple paths within a component. Moreover, the k-coreness is
  strongly correlated with the largest eigenvalue; we also performed
  regressions using the k-coreness instead of multi-path growth and
  the results remained basically unchanged.} We expect the probability
of an alliance to be negatively correlated with the average multi-path
growth. Indeed, low values of this variable indicate that at least one
of the potential partners (or both) is not in a component with a
strong multi-path growth, thus increasing the incentive to form an
alliance.\footnote{The largest eigenvalue of a component is the same
  for two potential partners if they are already part of the same
  component and increases with the number of links within the
  component. Moreover, the change in such eigenvalue decreases with
  the size of the component in many network structures. This implies a
  lower incentive to form an alliance if the two partners are already
  part of components with many independent multiple paths. See
  \citet{KoBaNaSchJEBO} and \citet{konig2012efficiency} for more
  details.}  For mixed dyads we use instead \textit{Incumbent
  multi-path growth}, that corresponds to the largest eigenvalue of
the connected component to which the potential incumbent partner
belongs. In a multiconnectivity based dynamics, the probability of
observing an alliance between an incumbent and an entrant can, on the
one hand, increase with incumbent multi-path growth, as new firms
benefit from attaching to a firm that has already access to many
paths. On the other hand, incumbents that have already access to many
paths may have less incentive to form an alliance with the entrant,
which is completely disconnected from the network, thus reducing the
probability of observing an alliance between the two.

\paragraph{Control variables.}
We control for a set of variables that can affect the formation of
alliances, both at the dyadic and at the aggregate level.

\medskip 

\textit{Time.} We employ time-fixed effect models, including dummy
variables for the year in which the dyads are observed. In this way we
control for unobserved effects present at the time an alliance is
formed.  Similar to \citet[][]{rosenkopf2008investigating}, we have
also tested a series of models including a time-trend
variable. However, all these models were characterized by a lower
goodness of fit than time-fixed effect models.

\medskip

\textit{Repeated alliances.} To control for repeated ties we include
an integer variable, \textit{Past alliances}, expressing the number of
alliances between the two firms in the dyad until the observation
year. In addition, as suggested by \citet{rosenkopf2008investigating},
we also consider the square value of such a variable in order to
control for possible non-linear effects.

\medskip

\textit{Sectoral alliance activity.} To control for sectoral trends in
alliances \citep[see e.g.][]{powell2005network} we compute the number
of alliances formed in the industrial sectors of the two firms in the
year preceding our observation. We then average the values for the two
firms in the observed dyad, obtaining the variable \textit{Sector
  alliances}.

\medskip 
We summarize the nomenclature and the meaning of all our
variables in Table \ref{table:model_variables}, and their correlations
and basic statistics in Table \ref{table:correlations}.

\renewcommand{\arraystretch}{1.0}
\begin{table}[h!]
\scriptsize
\begin{center}
\begin{tabular}{p{3.9cm} p{2.3cm} p{8.6cm} }
\hline \hline
  & \textbf{Type} & \textbf{Meaning} \\ \\
\hline
\textbf{Dependent variable} & & \\
\hline \\
\textit{Alliance formation} & binary & Formation of a link between the
two considered firms in the considered year. \\ \\

\hline
\multicolumn{2}{l}{\textbf{Independent variables}}  & \\
\hline \\

\multicolumn{2}{l}{\textbf{1. Accumulative advantage}} & \\
\hline
\textit{Joint centrality} & positive real & Logarithm of the arithmetic mean of the
degree centrality of the two firms (for incumbent dyads). \\ 
\textit{Incumbent centrality} & positive real  & Logarithm of the degree
centrality of the incumbent firm (for mixed dyads). \\ \\

\multicolumn{2}{l}{\textbf{2. Structural homophily and diversity}}  & \\
\hline
\textit{Same nation} & binary & 1 if the two firms are registered in the same nation. \\
\textit{Same SIC}   & binary & 1 if the considered firms have the same SIC code. \\
\textit{Technological distance} & positive real & Distance between the two firms in a technology space (measured through patent similarity). \\
\textit{Centrality inequality} & positive real & Logarithm of the ratio of the
degree centralities -- the largest divided by the smallest -- of
the two firms (only for incumbent dyads). \\
\textit{Inverse path length} & positive real & Inverse of the length
of the shortest network path connecting the two firms (only for incumbent dyads). \\
\textit{Common neighbors} & positive integer &  Number of the alliance
partners that the two firms have in common (only for incumbent
dyads). \\ \\

\multicolumn{2}{l}{\textbf{3. Multiconnectivity}} & \\
\hline
\textit{Average multi-path growth} & real & Arithmetic mean of the largest eigenvalues of the connected components to which the two firms belong (for incumbent dyads)
 \\
\textit{Incumbent multi-path growth} & real & Largest eigenvalue of the connected
components to which the incumbent firm belongs (for mixed dyads). \\ \\

\multicolumn{2}{l}{\textbf{Control variables}}  & \\
\hline
\textit{Year dummies} & binary & 23 dummy variables (the observation period consists of 24 years) for the time-fixed effect models. \\
\textit{Past alliances}  & positive integer & Number of alliances
established between the two firms in all previous years (only for
incumbent dyads). \\
\textit{(Past alliances)$\mathit{^2}$} & positive integer & Square of
the variable \textit{Past alliances} (only for incumbent dyads). \\
\textit{Sector alliances} & positive real & Arithmetic mean of the number of alliances in the industrial sectors of the two firms, in the year preceding the observation. \\
\hline \hline
\end{tabular}
\caption[]{Nomenclature, type and meaning of all the variables employed in our econometric model. The observation unit consists of a firm dyad, i.e. every potential pair of firms in the network.}
\label{table:model_variables}
\end{center}
\end{table}

\renewcommand{\arraystretch}{1.1}
\begin{table}[h!]
\tiny
\begin{tabular}{p{3.5cm}|rrrr|rrrrrrrrrrrrrrr}

\multicolumn{10}{l}{Risk set A: both firms in the dyad are incumbents (num.obs.= 766,733)} \cr
Variable & Mean & S.D. & Min. & Max. & 1 & 2 & 3 & 4 & 5 & 6 & 7 & 8 & 9 & 10 & 11 & 12 \cr 
\hline
1. \textit{Alliance formation}             & 0.00 & 0.05 & 0 & 1 & -- & -- & -- & -- & -- & -- & -- & -- & -- & -- & -- & -- \cr 
2. \textit{Joint centrality}          &  1.14 & 0.80 & 0 &  4.28 & 0.05 & -- & -- & -- & -- & -- & -- & -- & -- & -- & -- & -- \cr 
3. \textit{Same nation}                   &  0.42 & 0.49 & 0 & 1 & 0.01 &   0.03 & -- & -- & -- & -- & -- & -- & -- & -- & -- & -- \cr 
4. \textit{Same SIC}                       & 0.15 & 0.36 & 0 & 1 & 0.03 &  -0.08 &   0.02 & -- & -- & -- & -- & -- & -- & -- & -- & -- \cr 
5. \textit{Technological distance}      & 0.75 & 0.30 & 0 & 1.41 & -0.06 &  -0.06 &   0.04 &  -0.37 & -- & -- & -- & -- & -- & -- & -- & -- \cr 
6. \textit{Centrality inequality}      & 0.98 & 0.82 & 0 &  4.37 & 0.01 &   0.65 &   0.01 &  -0.05 &  -0.02 & -- & -- & -- & -- & -- & -- & -- \cr 
7. \textit{Inverse path length}            & 0.18 & 0.16 & 0 & 1 & 0.09 &   0.46 &   0.08 &   0.11 &  -0.21 &   0.14 & -- & -- & -- & -- & -- & -- \cr 
8. \textit{Common neighbors}              & 0.11 & 0.69 & 0 & 30 & 0.13 &   0.27 &   0.04 &   0.06 &  -0.14 &   0.01 &   0.49 & -- & -- & -- & -- & -- \cr 
9. \textit{Average multi-path growth}  & 13.41 & 5.49 & 1 & 19.1 & 0.00 &   0.38 &   0.09 &  -0.08 &   0.06 &   0.17 &   0.53 &   0.08 & -- & -- & -- & -- \cr 
10. \textit{Time}                       & 10.75 & 3.44 & 0 &  23 & -0.02 &  -0.17 &  -0.08 &   0.08 &  -0.06 &  -0.06 &  -0.21 &  -0.01 &  -0.37 & -- & -- & -- \cr 
11. \textit{Past alliances}                & 0.01 & 0.15 & 0 & 15 & 0.00 &   0.00 &   0.00 &   0.00 &   0.00 &   0.00 &   0.01 &   0.00 &   0.01 &   0.00 & -- & -- \cr 
12. \textit{(Past alliances)$\mathit{^2}$} & 0.02 & 1.02 & 0 & 225 & 0.00 &   0.00 &   0.00 &   0.00 &   0.00 &   0.00 &   0.00 &   0.00 &   0.00 &   0.00 &   0.71 & -- \cr 
13. \textit{Sector alliances}            & 80.80 & 56.07 & 0 & 286 & 0.01 &   0.12 &   0.12 &   0.32 &   0.11 &   0.04 &   0.15 &   0.02 &   0.27 &  -0.23 &   0.00 &   0.00 \cr 
\hline
\end{tabular}

\vspace{8pt}
\begin{tabular}{p{3.5cm}|rrrr|rrrrrrr}
\multicolumn{10}{l}{Risk set B: one firm in the dyad is a new-entrant (num.obs.= 1,432,811)} \cr
Variable & Mean & S.D. & Min. & Max. & 1 & 2 & 3 & 4 & 5 & 6 & 7 \cr 
\hline
1. \textit{Alliance formation}             & 0.00 & 0.02 & 0 & 1 & -- & -- & -- & -- & -- & -- & -- \cr 
2. \textit{Incumbent centrality}          &  0.76 & 0.86 & 0 & 4.37 & 0.02 & -- & -- & -- & -- & -- & -- \cr 
3. \textit{Same nation}                   &  0.36 & 0.48 & 0 & 1 & 0.01 &   0.00 & -- & -- & -- & -- & -- \cr 
4. \textit{Same SIC}                       & 0.14 & 0.34 & 0 & 1 & 0.02 &  -0.04 &   0.00 & -- & -- & -- & -- \cr 
5. \textit{Technological distance}      & 0.78 & 0.30 & 0 & 1.41 & -0.02 &  -0.04 &   0.03 &  -0.31 & -- & -- & -- \cr 
6. \textit{Incumbent multi-path growth}  & 9.29 & 6.56 & 1 & 19.1 & 0.00 &   0.41 &   0.04 &  -0.04 &   0.01 & -- & -- \cr 
7. \textit{Time}                       & 12.46 & 5.16 & 0 &  23 & -0.01 &  -0.08 &  -0.07 &   0.05 &  -0.01 &  -0.26 & -- \cr 
8. \textit{Sector alliances}            & 56.82 & 48.05 & 0 & 286 & 0.01 &   0.10 &   0.04 &   0.37 &   0.10 &   0.25 &  -0.01 \cr 
\hline
\end{tabular}

\caption[]{Descriptive statistics and correlations of all the variables employed in our regression models, for incumbent dyads (risk set A) and mixed dyads (risk set B).}
\label{table:correlations}
\end{table}
\renewcommand{\arraystretch}{1.0}

\subsection{Statistical methodology and regressions results} 
\label{sec:regressions-results}

Given the binary nature of our dependent variable, we use binomial
regressions for all our models. The model fitting is done through
Maximum Likelihood Estimation (MLE). Following the approach proposed
by \citet{nesta_complementary}, we employ a complementary log-log link
function, particularly suited for an abundance of rare events.
Differently from the logit and the probit link functions, the
complementary log-log function is asymmetrical, and thus frequently
used when the probability of the examined event is very large or very
small.\footnote{However, even though the number of successes, or
  ``ones'', in our dependent variable is low compared to the total
  number of observations (0.12 \%), it is high enough in absolute value
  -- we have a total of 2,597 link formation events in our sample -- thus
  making our results statistically significant \citep[see][for a more
  detailed discussion on the topic]{nesta_complementary,
    KingZeng2001_logistic_rare_events}.}  If we call the binary
dependent variable \textit{Alliance formation} $A$, adopting the
definitions of \citet{nesta_complementary}, the dependence of $A$ on
the independent variables $\mathbf{X}$ can be written as:
\begin{equation}
  \mathrm{Prob}(A=1|\mathbf{X}) = 1- \mathrm{exp}[- \mathrm{exp} (\mathbf{X} \boldsymbol{\beta}) ]
\end{equation}
where $\boldsymbol{\beta}$ is the vector of coefficients. We report
their estimates and their confidence intervals for our different
models, risk sets and time periods in Table \ref{table:models_incumbents}. The
goodness of fit of each models in each risk set is expressed through
the Akaike Information Criterion (AIC), and the Likelihood ratio with
respect to the baseline model, which is the one including only the
variables related to accumulative advantage.

\begin{table}
\scriptsize
\begin{center}
\begin{tabular}{l D{.}{.}{6.5}@{} D{.}{.}{6.5}@{} D{.}{.}{6.5}@{} D{.}{.}{6.5}@{} D{.}{.}{6.5}@{} D{.}{.}{6.5}@{} }
\hline
\multicolumn{1}{l}{\textbf{Risk set A --}} & \multicolumn{3}{c}{Rise phase (1986--1997)} & \multicolumn{3}{c}{Fall phase (1998--2009)}  \\
\multicolumn{1}{l}{\textbf{Incumbent dyads}}   & \multicolumn{1}{r}{Model 1A} & \multicolumn{1}{r}{Model 2A} & \multicolumn{1}{r}{Model 3A} & \multicolumn{1}{r}{Model 4A} & \multicolumn{1}{r}{Model 5A} & \multicolumn{1}{r}{Model 6A} \\
\hline
\\
(Intercept)                  & -4.766^{**} & -4.375^{**} & -4.337^{**} & -9.577^{**} & -7.168^{**} & -7.127^{**} \\
                             & (0.998)     & (0.978)     & (0.979)     & (0.264)     & (0.288)     & (0.494)     \\
\textit{Joint centrality}    & 1.552^{**}  & 1.016^{**}  & 1.043^{**}  & 1.402^{**}  & 1.003^{**}  & 1.006^{**}  \\
                             & (0.036)     & (0.048)     & (0.049)     & (0.077)     & (0.105)     & (0.109)     \\
\textit{Same nation}         &             & 0.368^{**}  & 0.370^{**}  &             & 0.039       & 0.039       \\
                             &             & (0.054)     & (0.054)     &             & (0.120)     & (0.120)     \\
\textit{Same SIC}            &             & 0.659^{**}  & 0.657^{**}  &             & 0.353^{*}   & 0.352^{*}   \\
                             &             & (0.066)     & (0.065)     &             & (0.146)     & (0.146)     \\
\textit{Technological distance}   &        & -2.315^{**} & -2.305^{**} &             & -3.439^{**} & -3.440^{**} \\
                             &             & (0.111)     & (0.111)     &             & (0.272)     & (0.272)     \\
\textit{Centrality inequality}   &         & -0.099^{**} & -0.114^{**} &             & -0.412^{**} & -0.412^{**} \\
                             &             & (0.036)     & (0.036)     &             & (0.087)     & (0.087)     \\
\textit{Inverse path length} &             & 2.362^{**}  & 2.443^{**}  &             & 2.134^{**}  & 2.139^{**}  \\
                             &             & (0.115)     & (0.114)     &             & (0.223)     & (0.228)     \\
\textit{Common neighbors}    &             & 0.016       & 0.010       &             & -0.040      & -0.041      \\
                             &             & (0.010)     & (0.011)     &             & (0.029)     & (0.029)     \\
\textit{Average multi-path growth} &       &             & -0.080^{**} &             &             & -0.003      \\
                             &             &             & (0.019)     &             &             & (0.030)     \\
\textit{Past alliances}      & 0.250       & 0.135       & 0.126       & 0.602       & 0.518       & 0.517       \\
                             & (0.452)     & (0.414)     & (0.411)     & (0.709)     & (0.795)     & (0.796)     \\
\textit{(Past alliances)$\mathit{^2}$} & -0.155      & -0.088      & -0.084      & -0.138      & -0.116      & -0.116      \\
                             & (0.272)     & (0.231)     & (0.228)     & (0.321)     & (0.396)     & (0.397)     \\
\textit{Sector alliances}    & 0.001       & -0.001^{*}  & -0.001^{*}  & -0.001      & -0.001      & -0.001      \\
                             & (0.001)     & (0.001)     & (0.001)     & (0.002)     & (0.002)     & (0.002)     \\
Year dummies                 & \multicolumn{1}{c}{(yes)} & \multicolumn{1}{c}{(yes)}& \multicolumn{1}{c}{(yes)}& \multicolumn{1}{c}{(yes)}& \multicolumn{1}{c}{(yes)}& \multicolumn{1}{c}{(yes)} \\
                             &             &             &             &              &             &       \\
\hline
AIC                          & 17712.974   & 15404.617   & 15390.905   & 4359.410    & 3741.857    & 3743.846    \\
BIC                          & 17891.093   & 15649.530   & 15646.951   & 4527.007    & 3972.303    & 3984.767    \\
Log Likelihood               & -8840.487   & -7680.308   & -7672.452   & -2163.705   & -1848.928   & -1848.923   \\
Likelihood ratio test        & 0           & 2320.358    & 2336.070    & 0           & 629.554     & 629.564     \\
Deviance                     & 17680.974   & 15360.617   & 15344.905   & 4327.410    & 3697.857    & 3697.846    \\
Num. obs.                    & 505067      & 505067      & 505067      & 261666      & 261666      & 261666      \\
\hline
\multicolumn{7}{l}{\scriptsize{$^{**}p<0.01$, $^*p<0.05$}}
\end{tabular}
\caption{Estimates of the model coefficients on the incumbent dyads (standard error in parentheses). Models 1A and 4A test the effect of accumulative advantage only; models 2A and 5A test the effect of accumulative advantage and structural homophily (diversity); models 3A and 6A test the effect of accumulative advantage, structural homophily (diversity) and multiconnectivity. Models 1A, 2A and 3A are related to the rise phase (1986-1997) of the R\&D network, while models 4A, 5A and 6A are related to its fall phase (1998-2009). The likelihood ratio test of the models 2A and 3A is computed with respect to model 1A; the likelihood ratio test of the models 5A and 6A is computed with respect to model 4A.}
\label{table:models_incumbents}
\end{center}
\end{table}

\begin{table}
\scriptsize
\begin{center}
\begin{tabular}{l D{.}{.}{6.5}@{} D{.}{.}{6.5}@{} D{.}{.}{6.5}@{} D{.}{.}{6.5}@{} D{.}{.}{6.5}@{} D{.}{.}{6.5}@{} }
\hline
\multicolumn{1}{l}{\textbf{Risk set B --}} & \multicolumn{3}{c}{Rise phase (1986--1997)} & \multicolumn{3}{c}{Fall phase (1998--2009)}  \\
\multicolumn{1}{l}{\textbf{Mixed dyads}}   & \multicolumn{1}{r}{Model 1B} & \multicolumn{1}{r}{Model 2B} & \multicolumn{1}{r}{Model 3B} & \multicolumn{1}{r}{Model 4B} & \multicolumn{1}{r}{Model 5B} & \multicolumn{1}{r}{Model 6B} \\
\hline
\\
(Intercept)                  & -8.641^{**} & -7.874^{**} & -7.824^{**} & -9.934^{**} & -7.510^{**} & -7.106^{**} \\
                             & (1.000)     & (1.007)     & (1.007)     & (0.285)     & (0.311)     & (0.403)     \\
\textit{Incumbent centrality} & 0.759^{**} & 0.717^{**}  & 0.753^{**}  & 0.673^{**}  & 0.604^{**}  & 0.654^{**}  \\
                             & (0.044)     & (0.043)     & (0.045)     & (0.072)     & (0.072)     & (0.080)     \\
\textit{Same nation}         &             & 0.915^{**}  & 0.921^{**}  &             & 0.120       & 0.121       \\
                             &             & (0.084)     & (0.084)     &             & (0.140)     & (0.140)     \\
\textit{Same SIC}            &             & 1.203^{**}  & 1.204^{**}  &             & 1.430^{**}  & 1.427^{**}  \\
                             &             & (0.105)     & (0.105)     &             & (0.179)     & (0.179)     \\
\textit{Technology distance} &             & -2.316^{**} & -2.314^{**} &             & -3.140^{**} & -3.152^{**} \\
                             &             & (0.153)     & (0.153)     &             & (0.285)     & (0.286)     \\
\textit{Incumbent multi-path growth} &     &             & -0.051^{**} &             &             & -0.032      \\
                             &             &             & (0.017)     &             &             & (0.021)     \\
\textit{Sector alliances}    & 0.001       & -0.003^{**} & -0.003^{**} & 0.010^{**}  & -0.002      & -0.002      \\
                             & (0.001)     & (0.001)     & (0.001)     & (0.002)     & (0.002)     & (0.002)     \\
Year dummies                 & \multicolumn{1}{c}{(yes)} & \multicolumn{1}{c}{(yes)}& \multicolumn{1}{c}{(yes)}& \multicolumn{1}{c}{(yes)}& \multicolumn{1}{c}{(yes)}& \multicolumn{1}{c}{(yes)} \\
                             &             &             &             &             &             &             \\
\hline
AIC                          & 8927.428    & 8239.980    & 8234.192    & 4004.853    & 3647.701    & 3647.537    \\
BIC                          & 9085.523    & 8431.951    & 8437.456    & 4167.831    & 3845.603    & 3857.080    \\
Log Likelihood               & -4449.714   & -4102.990   & -4099.096   & -1988.427   & -1806.850   & -1805.768   \\
Likelihood ratio test        & 0           & 693.448     & 701.236     & 0           & 363.154     & 365.318     \\
Deviance                     & 8899.428    & 8205.980    & 8198.192    & 3976.853    & 3613.701    & 3611.537    \\
Num. obs.                    & 592703      & 592703      & 592703      & 840108      & 840108      & 840108      \\
\hline
\multicolumn{7}{l}{\scriptsize{$^{**}p<0.01$, $^*p<0.05$}}
\end{tabular}
\caption{Estimates of the model coefficients on the mixed dyads (standard error in parentheses). Models 1B and 4B test the effect of accumulative advantage only; models 2B and 5B test the effect of accumulative advantage and structural homophily (diversity); models 3B and 6B test the effect of accumulative advantage, structural homophily (diversity) and multiconnectivity. Models 1B, 2B and 3B are related to the rise phase (1986-1997) of the R\&D network, while models 4B, 5B and 6B are related to its fall phase (1998-2009). The likelihood ratio test of the models 2B and 3B is computed with respect to model 1B; the likelihood ratio test of the models 5B and 6B is computed with respect to model 4B.}
\label{table:models_newcomers}
\end{center}
\end{table}

\paragraph{Results.} 
By comparing the AIC scores of the different model sets, we find that the variables related to the accumulative advantage mechanism alone are already sufficient to achieve a fairly good predictive power, both in the rise phase (models 1A and 1B) and in the fall one (models 4A and 4B).
In particular, the variable \textit{Joint centrality} for the incumbent dyads displays always positive and significant coefficients in all models (models 2A,  3A, 5A and 6A), thus validating the hypothesis that alliances are more likely to be formed between more central firms -- in particular, firms with more distinct partners. The same finding holds for the mixed dyads (see the variable \textit{Incumbent centrality} in models 2B, 3B, 5B and 6B), proving that new-entrant firms are more likely to become part of the R\&D network by attaching to the most central incumbents.

\medskip
The inclusion of the variables related to the structural homophily and
diversity of the dyad  yields a better predictive power
than the baseline model, as indicated both by the likelihood test
ratios and the AIC scores (compare model 2A with 1A, 5A with 4A, 2B
with 1B, and 5B with 4B). Remarkably, we obtain robust results across
models and risk sets. The variable \textit{Same SIC} exhibits a
positive and significant effect for both incumbent and mixed dyads, in
both the rise and the fall phases, indicating that firms tend to form intra-sectoral rather than inter-sectoral R\&D alliances.
In contrast, the variable \textit{Same nation} exhibits a positive and significant
effect in the rise phase -- both for incumbent and mixed dyads -- but not in the fall phase (where it is not significant), indicating that firms lose the tendency to form geographically close alliances during the ``fall'' of R\&D networks.
The variable \textit{Technological distance} exhibits a negative and
significant effect for both incumbent and mixed dyads, in both the
rise and the fall phases, bringing additional support to the argument
of the importance of absorptive capacity considerations in the
formation of R\&D alliances
\citep[see][]{cohen1990absorptive}. To sum up, geographical, sectoral
and technological \textit{similarities} are positively correlated with the formation of a
new alliance.

\medskip Next, the variable \textit{Centrality inequality}, computed
only on incumbent dyads, exhibits a negative and significant effect in
both the rise and the fall phases. This means that the dyads in which
one firm is much more central than the other one are less likely to
form an alliance; on the contrary, firms with similar degree
centrality are more likely to form an alliance.  This finding is in
agreement with the positive assortativity coefficient that we have
reported for the R\&D networks in Section
\ref{sec:homoph-hetero-alli-behav}, showing that firms with similar
degree centrality are more likely to be connected.  Furthermore, the
variable \textit{Inverse path length} exhibits a positive and
significant effect across models and time periods, stressing the
importance of network-endogenous mechanisms for selecting new alliance
partners \citep[see][]{rosenkopf2008investigating}. We find that firms
which are already connected by a path in the network are more likely
to form an alliance. Moreover, the shorter this path is, the more
likely the alliance is formed.  Finally, the variable \textit{Common
  neighbors} is not significant at any time for any type of dyads,
indicating that structural homophily in terms of partners similarity is
not an effective predictor of the formation of a new R\&D alliance.

\medskip
Regarding \textit{multiconnectivity} we find that, for incumbent
dyads, the variable \textit{Average multi-path growth} shows a
negative and significant effect in the rise phase, indicating that
firms embedded in a network component with a small eigenvalue are more
likely to form an alliance. Likewise, in mixed dyads, the new-entrant
firm tends to connect to an incumbent that is located in a component
with a small eigenvalue, as shown by the coefficient of the variable
\textit{Incumbent multi-path growth}. This result 
supports the hypothesis of a network-growth mechanism in which firms
try to increase both the number of new partners and the number of new
independent pathways to which they can get access when they form
new alliances. The fact that firms with a small
multi-path growth score are more likely to form alliances indicates
that they actually attempt to increase this score by means of new
network ties. However, differently from all the other modes of
alliance formation, multiconnectivity does not matter in the fall
phase of the network, as shown by the fact the variable
\textit{Average multi-path growth} 
is not significant in the period 1998-2009, both for incumbents and
mixed dyads.

Finally, for incumbent dyads, the control variables \textit{Past alliances} and its square exhibit, respectively, a positive and a negative coefficient. In principle, this is consistent with the prediction of the non-linear effect of repeated alliances on the formation of a new link \citep[see][]{rosenkopf2008investigating}. However, these coefficients are never significant in our regressions, in any risk set or model variant, thus preventing us from drawing any conclusion.
the control variable \textit{Sector alliances} exhibits a negative effect, in the rise phase of the R\&D network, which is more significant for mixed dyads rather then incumbent dyads. This confirms the presence of competition and saturation effects during the fast growth of the R\&D network (1986-1997), which make the establishment of a new R\&D alliance less likely in sectors already showing a high alliance activity in the previous year. Such effects are stronger when a new-entrant firm is involved in the alliance, clearly showing the presence of an entry barrier to the R\&D network.

\medskip

The above finding can be summarized as follows: The rise of the global
network of R\&D alliances has largely been driven by accumulative
advantage factors as well as structural \textit{homophily} -- rather
than \textit{diversity} -- considerations. In their search for R\&D
partners, firms were influenced both by the need of establishing
connections with more central partners in the network, as well as by
similarity in technonological and (at least in the rise phase) spatial
proximity. In turn, these trends were reinforced by alliance formation
strategies based on multiconnectivity aimed at increasing the number
of pathways through which other parters could be reached. The last driver is
compatible with the emergence of a large and densely connected
network, observed at the end of the rise phase. During the fall phase,
instead, alliance formation was still driven by accumulative advantage
and structural homophily, but not by multiconnectivity. It follows
than in such a phase, firms were still looking for more central and more similar partners,
but disregarding the possibility of getting indirect access to other
firms in the network. In its turn, this explains the disappearance of
bridging ties and the fragmentation of the R\&D network.

\section{Concluding remarks}
\label{sec:concluding-remarks}

In this study we have empirically investigated the dynamics and the properties of a set of
global inter-firm networks of R\&D alliances, from 1986 to 2009. We
considered both a ``pooled'' R\&D network, i.e.
a network where alliances are considered independently of the
sectors of the partners, as well as ten sectoral R\&D
networks for the largest industrial sectors represented in our
dataset. On the grounds of this structural analysis we have then
investigated via binomial regressions the drivers of alliance
formation, by testing the relevance of some mechanisms
that have so far been proposed in the empirical literature. 

Our results provide strong support to the claim that several
properties of R\&D networks are not only robust across several
manufacturing sectors, but also invariant across different scales of
aggregation. In other words, they are stable if one considers the
pooled R\&D network or the sectoral networks. These properties do not
only relate to basic network characteristics like size, density,
degree distributions. They also include more complex features
concerning the organization of the network components, such as the
degree of structural homophily (captured by the assortativity
coefficient) and the presence of small worlds. Our results generalize
previous findings in the literature, that have been limited to the
analysis of few sectors.  They also provide empirical support to the
idea that the process of alliance formation can be analyzed
independently of the sectoral specificities of the firms involved in
R\&D alliances and -- similar to many previous empirical studies
\citep[e.g.][]{powell2005network,gulati2012:_rise_and_fall_of_small_world,rosenkopf2008investigating} --
it can be described in terms of simple rules. Such rules determine the
probability of forming alliances on the basis of the firms' position in
the network and on the properties of the existing network structures. 
In addition, the finding that networks
display a core-periphery architecture provides an important
refinement with respect to the exisiting knowledge on network
properties, as this feature is able to jointly explain both fat-tailed
degree distributions and the presence of small worlds, which have received
great attention in previous empirical studies.

Furthermore,  we find that the last three decades have
witnessed a rise-and-fall of R\&D networks at all scales of aggregation. Such a
rise-and-fall dynamics has been previously emphasized only with respect to the
presence of small worlds in the computer industry
\citep{gulati2012:_rise_and_fall_of_small_world}. We show that it is
instead a general property of the network dynamics, involving many
network indicators (and not only the presence of small
worlds). Our regression results indicate that a structural break in
some drivers of alliance formation is likely to be at the basis of the
above life cycle 
in R\&D networks. In particular, we find that
both in the rise and in the fall phase the alliance formation has been driven by both ``accumulative advantage'' (the search for more central partners in the network) and ``structural homophily'' (the search for more similar partners, in terms of industrial sector, technology and geographical proximity).
In contrast, firms have formed alliances to expand the number of indirect paths to other firms in the network \citep[the ``multiconnectivity'' driver, see][]{powell2005network} in the rise phase only. We have, indeed, detected a loss of significance of the latter alliance driver in the fall phase, thus providing an explanation for the observed disappearance of bridging ties and the subsequent fragmentation of the network.

Our work could be extended in at least two ways. First, one could
improve the regression analyses with measures related to the industry
demand and market structure, to check how the characteristics of
within-industry competition may affect the formation of intra- and
inter-industry R\&D collaboration.  Second, building on the empirical
evidence one could develop an
agent-based model to reproduce the emergent network properties
through a bottom-up approach.  In particular, such a new agent-based
model should be able to predict at the same time all the features that
we have empirically observed in the R\&D networks, namely degree
distributions, assortativity, presence of small-world and nested
architectures. The goal is to eventually unveil the complex
interdependencies and mutual feedbacks between the emergent network
structures and the individual firms' decisions.

\end{spacing}

\section*{Acknowledgements}
We thank Lionel Nesta, Claudio J. Tessone, Moritz M\"{u}ller,  Stefano Battiston,
Giorgio Fagiolo, Nobuyuki Hanaki, Sanjeev Goyal, Pier-Paolo Saviotti,
Pier-Paolo Andriani and Bulat Sanditov, for useful comments and discussions. In addition,
we thank the participants to the European Conference on Complex Systems 2012, in Brussels
(Belgium), to the Latsis Symposium 2012, in Z\"{u}rich
(Switzerland), to the EMAEE 2013 conference, in
Sophia-Antipolis (France), and to the WEHIA 2013
conference, in Reykjav\'{i}k (Iceland). M.~V.~T, acknowledges
financial support from the Swiss National Science Foundation (SNF)
through grant 100014\_126865. A.~G. and F.~S. acknowledge financial support by
the EU-FET project MULTIPLEX 317532. M.N. acknowledges financial
support by the EU H2020 project ISIGrowth, under grant agreement 649186.

\section*{Appendix}
\label{appendix}

\paragraph{Degree distributions.}
In our network representation, we count multiple R\&D alliances
between the same two firms as one, and we count all the firms
participating in the same consortia as distinct partners. Furthermore,
similarly to Section \ref{sec:basic-netw-prop}, the whole observation
period is divided into six sub-periods lasting 4 years. All the
measures we present are computed by aggregating firm degree data
relative to the same sub-period. Fig.~\ref{fig:degree_distr} shows the
degree distributions of the pooled R\&D network in the six analyzed
sub-periods. More precisely, given each degree distribution, we report
its \textit{complementary cumulative distribution function} $P(x)$,
defined as the fraction of nodes having degree greater than or equal
to $x$:
\begin{equation} \label{eq:cdf} P(x)= \int_x^\infty p(x')
  \mathrm{d}x'.
\end{equation}
where $p(x')$ is the \textit{probability density function}, defining
the fraction of nodes in the network with degree $x$. The
complementary cumulative distribution function is more robust than the
probability density function against fluctuations due to finite sample
sizes (particularly in the tail).

\paragraph{Hill Estimator.}
Let us assume that we have a network with $N$ nodes. If $N$ is the
number of observations (in our case, the number of node degrees that
we measure in the R\&D network) and $t$ is the number of tail
observations ($t \le N$), the inverse of the Hill Estimator (HE) is
defined as:
\begin{equation} \label{eq:Hill_estimator} \mathrm{HE}^{-1} = t^{-1}
  \sum_{i=1}^{t} \left[ \log(x_{i}) - \log(x_{min}) \right],
\end{equation}
where $x_{min}$ represents the beginning of the tail and $x_{i}$, $i=1
\dots t$ are the tail observations, i.e. the degree values such that
$x_{i} \ge x_{min}$. The smaller the HE value, the ``heavier'' the
tail of the degree distribution is. In particular, the degree
distributions of most biological, social and economic systems display
values of the HE between 2 and 4 \citep[see][]{clauset2009power}.

\begin{table}[h!]
  \scriptsize
  \centering
  \begin{tabular*}{0.99\linewidth}{@{\extracolsep{\fill}}l*{6}{@{ }r}}
    \hline
    & \textbf{1986-1989} & \textbf{1990-1993} & \textbf{1994-1997} & \textbf{1998-2001} & \textbf{2002-2005} & \textbf{2006-2009} \cr
    \hline

    \textbf{Pooled} & 0.38 & 0.54 & 0.67 & 0.62 & 0.46 & 0.62 \cr 
    Pharmaceuticals & 0.42 & 0.42 & 0.42 & 0.50 & 0.46 & 0.46 \cr 
    Computer Software & 0.33 & 0.67 & 0.42 & 0.50 & 0.50 & 0.46 \cr 
    Electronic Components & 0.54 & 0.42 & 0.54 & 0.50 & 0.50 & 0.46 \cr 
    Computer Hardware & 0.62 & 0.62 & 0.29 & 0.21 & 0.38 & 0.46 \cr 
    Medical Supplies & - & 0.42 & 0.29 & 0.58 & 0.58 & 0.67 \cr 
    Communications Equipment & 0.54 & 0.54 & 0.50 & 0.54 & 0.46 & 0.46 \cr 
    Laboratory Apparatus & 0.42 & 0.21 & 0.46 & 0.50 & 0.38 & 0.38 \cr 
    Motor Vehicles & 0.54 & 0.46 & 0.21 & 0.58 & 0.67 & 0.46 \cr 
    Inorganic Chemicals & 0.50 & 0.46 & 0.46 & 0.58 & 0.38 & 0.42 \cr 
    Aircrafts and parts & 0.38 & 0.38 & 0.42 & 0.42 & 0.67 & 0.50 \cr 

    \hline \hline
  \end{tabular*}
  \caption{P-values of the Kolmogorov-Smirnov tests for the Hill Estimators (HE) of the degree distributions in the pooled
    and the sectoral R\&D networks. The null hypothesis is that the empirical data are drawn from a distribution having the fitted HE value. \textit{Note}: small p-values (less than 0.05) would indicate that the null hypothesis has to be rejected, whilst large p-values (greater than 0.05) indicate that the null hypothesis cannot be rejected.}
  \label{table:he_pvalues}
\end{table}

In Table \ref{table:he_pvalues} we report the p-values of the Kolmogorov-Smirnov tests performed on each R\&D network and time period -- cf. the fitted HE values reported in Section \ref{sec:homoph-hetero-alli-behav}, Table \ref{table:he}. Small p-values (less than 0.05) would indicate that the test rejects the hypothesis that the original data are drawn from a distribution having the fitted HE value. In \textit{all} cases, we obtain values greater than 0.05, confirming the significance of all reported HE values.

\paragraph{Assortativity mixing coefficient.}
To investigate assortativity-disassortativity in our R\&D networks, we
use the assortativity mixing coefficient $r$ proposed by
\citet{newman02:_assor_mixin_in_networ}. This quantity, as described
by Eq. \ref{eq:assortativity_coeff}, is the Pearson correlation
coefficient of the degrees at both ends of all links in the network:
\begin{equation} \label{eq:assortativity_coeff} r = \frac{4M^{-1}
    \sum_{i}{j_{i}k_{i}} - [M^{-1} \sum_{i}{(j_{i}+k_{i})}]^{2}
  }{2M^{-1} \sum_{i}{(j_{i}^{2}+k_{i}^{2})} - [M^{-1}
    \sum_{i}{(j_{i}+k_{i})}]^{2} },
\end{equation}
where $j_{i}$, $k_{i}$ are the degrees of the firms at the ends of the
$i$-th link, with $i=1,...,M$.  The coefficient $r$ ranges between
$-1$ for a totally disassortative network to $1$ for a totally
assortative network; a network in which links are formed randomly
would exhibit $r=0$.  For instance,
\citet{ramasco04:_self_organ_of_collab_networ} develop models wherein
agents establish links with most central actors in the network, and
show that such a mechanism gives rise to disassortative
networks. However, \citet{konig2010assortative} show that the same
mechanism of search for high centrality can give rise to assortative
networks if agents face limitations in the number of collaborations
they are able to maintain.

\paragraph{Small world coefficient.}
According to
\citet{strogatz98:_collective_dynamics_small_world_networks}, the
small world properties of a network have to be evaluated using a
corresponding random network as the baseline. If the examined network
is both large and sparse, i.e. $N \gg \bar{k}$, where $N$ is the
network size and $\bar{k}$ is the average degree, the basic
requirement for small world is satisfied. Under this assumption, the
values of clustering coefficient $C$ and average path length $L$ for
the baseline random network will tend to: $C_{R}=\bar{k}/N$ and
$L_{R}=\log(N)/\log(\bar{k})$. The small world quotient $Q_{SW}$ we
use for our analysis is defined as:

\begin{equation} \label{eq:sw_quotient} Q_{SW} = \frac{(C/L)}{(C_{R}/L_{R})}=
  \frac{(C/C_{R})}{(L/L_{R})}.
\end{equation}

In our study, the condition of sparse network is always fulfilled for
the pooled and the sectoral R\&D networks (the average degrees are
always smaller than 3 in our sample, and much smaller than the corresponding network
sizes). Some of the sectoral R\&D networks have relatively small sizes in the first
(1986-1989) and in the last (2006-2009) observation periods (as can be
seen from Table~\ref{table:size}), but in these cases they exhibit an
even smaller average degree $\bar{k}$, still validating the assumption
of sparse networks. When computing the observed to random ratios, a
small world network will show $C/C_{R} \gg 1$ and $L/L_{R} \simeq 1$,
which is the case for all the R\&D networks we analyze.

\paragraph{Nestedness coefficient.}
In order to quantify the extent to which a network displays
core-periphery nested structures, we use the \textit{BINMATNEST} algorithm, proposed by
\citet{rodriguez2006nestedness}. The algorithm uses the unweighted adjacency matrix
of the network to compute its nestedness score. The adjacency matrix
is rearranged in such a way that all the ``ones''
(existing links) are concentrated in the top-left side of the matrix,
and the ``zeros'' (missing links) in the bottom-right side. The algorithm
then computes the optimal theoretical isocline separating the ``ones'' from
the ``zeros'' and counts the number of holes in these regions of the
matrix -- i.e. how many ``zeros'' are in the region of the ``ones'',
and vice versa. The number of such holes is proportional to the
score computed by the algorithm. \textit{Note:} the lower the
score, the more nested the network is (and vice versa). The value
returned by the algorithm, $T_{n}$,
ranges from 0, for a totally nested network, to 100, for a completely
random (non-nested) network. Instead of directly using
the value generated by the algorithm, we use a normalized nestedness
coefficient $C'_{n}$, which we define as:
\begin{equation} \label{eq:nestedness} C'_{n} = \frac{100-T_{n}}{100},
\end{equation}
where $T_{n}$ is the nestedness score generated by the
\textit{BINMATNEST} algorithm. Our normalized nestedness coefficient
$C'_{n}$ has a more direct interpretation and
spans from 0 for a totally random (non-nested) network, to 1 for
a totally nested network.
An important remark is that the nestedness, by definition, can be computed only on connected networks. For this reason, we consider the largest connected component for each of the R\&D networks that we analyze.
The algorithm is also able to compute the p-values, by building
and analyzing 100 random networks having the same
size and density as the giant component of the network under examination. Here, the null hypothesis is that a random network with the same size and density as the network at hand exhibits the same nestedness coefficient. In Table \ref{table:nestedness_pvalues} we report the nestedness coefficients for all R\&D networks in all years -- not only the time averages (see Table \ref{table:nestedness}) -- with the corresponding p-values visually encoded as significance stars.

\renewcommand{\arraystretch}{1.1}
\begin{table}[h!]
  \tiny
  \centering
  \begin{tabular*}{1.0\linewidth}{@{\extracolsep{\fill}}l*{24}{@{ }l}}
    \hline
    & 1986 & 1987 & 1988 & 1989 & 1990 & 1991 & 1992 & 1993 & 1994 & 1995 & 1996 & 1997  \cr
    \hline

\textbf{Pooled} & $0.607^{*}$ & $0.660^{ }$ & $0.911^{*}$ & $0.985^{*}$ & $0.990^{**}$ & $0.996^{**}$ & $0.998^{***}$ & $0.998^{***}$ & $0.999^{***}$ & $0.999^{***}$ & $0.999^{***}$ & $0.998^{***}$ \cr 
Pharmaceuticals & $0.725^{ }$ & $0.744^{ }$ & $0.750^{ }$ & $0.605^{ }$ & $0.961^{*}$ & $0.989^{*}$ & $0.991^{**}$ & $0.992^{**}$ & $0.994^{**}$ & $0.994^{***}$ & $0.992^{**}$ & $0.990^{*}$ \cr 
Computer Software & $0.737^{ }$ & $0.660^{ }$ & $0.912^{ }$ & $0.985^{*}$ & $0.984^{*}$ & $0.989^{*}$ & $0.995^{**}$ & $0.996^{***}$ & $0.997^{***}$ & $0.997^{***}$ & $0.997^{***}$ & $0.996^{***}$ \cr 
Electronic Components & $0.737^{ }$ & $0.660^{ }$ & $0.805^{*}$ & $0.910^{*}$ & $0.957^{*}$ & $0.981^{*}$ & $0.989^{***}$ & $0.989^{***}$ & $0.992^{***}$ & $0.992^{***}$ & $0.992^{***}$ & $0.989^{***}$ \cr 
Computer Hardware & $0.623^{ }$ & $0.695^{ }$ & $0.911^{*}$ & $0.984^{*}$ & $0.983^{*}$ & $0.990^{*}$ & $0.995^{***}$ & $0.995^{***}$ & $0.996^{***}$ & $0.995^{***}$ & $0.995^{***}$ & $0.993^{***}$ \cr 
Medical Supplies & $0.623^{ }$ & $0.623^{ }$ & $0.737^{ }$ & $0.737^{ }$ & $0.680^{*}$ & $0.866^{*}$ & $0.885^{**}$ & $0.929^{**}$ & $0.954^{***}$ & $0.961^{***}$ & $0.945^{**}$ & $0.843^{**}$ \cr 
Communic. Equipment & $0.375^{ }$ & $0.375^{ }$ & $0.375^{ }$ & $0.729^{*}$ & $0.905^{*}$ & $0.956^{*}$ & $0.978^{***}$ & $0.982^{***}$ & $0.987^{***}$ & $0.990^{***}$ & $0.989^{***}$ & $0.986^{***}$ \cr 
Laboratory Apparatus & $0.737^{ }$ & $0.695^{ }$ & $0.769^{ }$ & $0.715^{ }$ & $0.720^{ }$ & $0.893^{*}$ & $0.943^{*}$ & $0.960^{***}$ & $0.967^{***}$ & $0.960^{***}$ & $0.931^{**}$ & $0.920^{**}$ \cr 
Motor Vehicles & $0.375^{ }$ & $0.695^{ }$ & $0.754^{ }$ & $0.695^{ }$ & $0.548^{ }$ & $0.916^{**}$ & $0.942^{***}$ & $0.935^{***}$ & $0.960^{***}$ & $0.959^{***}$ & $0.958^{***}$ & $0.918^{***}$ \cr 
Inorganic Chemicals & $0.355^{ }$ & $0.355^{ }$ & $0.769^{ }$ & $0.764^{ }$ & $0.905^{*}$ & $0.961^{*}$ & $0.970^{***}$ & $0.875^{***}$ & $0.972^{***}$ & $0.958^{***}$ & $0.910^{***}$ & $0.858^{*}$ \cr 
Aircrafts and parts & $0.607^{ }$ & $0.607^{ }$ & $0.769^{ }$ & $0.768^{ }$ & $0.625^{ }$ & $0.891^{*}$ & $0.930^{*}$ & $0.949^{***}$ & $0.932^{***}$ & $0.944^{***}$ & $0.823^{***}$ & $0.799^{**}$ \cr 

    \hline
    & 1998 & 1999 & 2000 & 2001 & 2002 & 2003 & 2004 & 2005 & 2006 & 2007 & 2008 & 2009 \cr
    \hline

\textbf{Pooled} & $0.997^{***}$ & $0.997^{***}$ & $0.995^{***}$ & $0.994^{***}$ & $0.995^{***}$ & $0.976^{***}$ & $0.988^{**}$ & $0.992^{***}$ & $0.993^{***}$ & $0.995^{***}$ & $0.995^{***}$ & $0.992^{**}$ \cr 
Pharmaceuticals & $0.991^{**}$ & $0.992^{**}$ & $0.991^{**}$ & $0.986^{*}$ & $0.989^{*}$ & $0.981^{*}$ & $0.989^{**}$ & $0.992^{**}$ & $0.993^{***}$ & $0.996^{**}$ & $0.995^{***}$ & $0.992^{*}$ \cr 
Computer Software & $0.992^{**}$ & $0.992^{**}$ & $0.965^{*}$ & $0.965^{*}$ & $0.956^{*}$ & $0.925^{*}$ & $0.933^{*}$ & $0.908^{*}$ & $0.833^{ }$ & $0.743^{ }$ & $0.823^{ }$ & $0.822^{ }$ \cr 
Electronic Components & $0.983^{**}$ & $0.981^{**}$ & $0.977^{*}$ & $0.978^{*}$ & $0.980^{*}$ & $0.963^{*}$ & $0.948^{*}$ & $0.910^{*}$ & $0.832^{*}$ & $0.807^{*}$ & $0.833^{*}$ & $0.806^{*}$ \cr 
Computer Hardware & $0.988^{**}$ & $0.986^{**}$ & $0.978^{*}$ & $0.976^{*}$ & $0.971^{ }$ & $0.942^{ }$ & $0.913^{*}$ & $0.858^{*}$ & $0.724^{ }$ & $0.713^{ }$ & $0.605^{ }$ & $0.605^{ }$ \cr 
Medical Supplies & $0.827^{**}$ & $0.676^{*}$ & $0.637^{*}$ & $0.663^{*}$ & $0.764^{ }$ & $0.769^{*}$ & $0.778^{*}$ & $0.688^{*}$ & $0.720^{*}$ & $0.833^{*}$ & $0.778^{*}$ & $0.769^{*}$ \cr 
Communic. Equipment & $0.975^{***}$ & $0.974^{**}$ & $0.960^{*}$ & $0.954^{*}$ & $0.963^{*}$ & $0.900^{*}$ & $0.908^{*}$ & $0.801^{*}$ & $0.799^{ }$ & $0.676^{*}$ & $0.432^{ }$ & $0.432^{*}$ \cr 
Laboratory Apparatus & $0.905^{*}$ & $0.935^{*}$ & $0.918^{*}$ & $0.884^{*}$ & $0.826^{*}$ & $0.830^{*}$ & $0.811^{*}$ & $0.725^{*}$ & $0.826^{*}$ & $0.696^{*}$ & $0.715^{*}$ & $0.695^{*}$ \cr 
Motor Vehicles & $0.904^{***}$ & $0.861^{***}$ & $0.896^{**}$ & $0.931^{**}$ & $0.944^{**}$ & $0.794^{*}$ & $0.786^{*}$ & $0.721^{*}$ & $0.746^{*}$ & $0.703^{**}$ & $0.423^{*}$ & $0.737^{*}$ \cr 
Inorganic Chemicals & $0.810^{*}$ & $0.725^{*}$ & $0.713^{*}$ & $0.793^{*}$ & $0.799^{*}$ & $0.749^{*}$ & $0.299^{*}$ & $0.737^{*}$ & $0.768^{*}$ & $0.767^{*}$ & $0.895^{*}$ & $0.791^{ }$ \cr 
Aircrafts and parts & $0.841^{*}$ & $0.792^{*}$ & $0.833^{*}$ & $0.838^{*}$ & $0.867^{*}$ & $0.679^{*}$ & $0.828^{ }$ & $0.828^{*}$ & $0.764^{*}$ & $0.764^{ }$ & $0.375^{ }$ & $0.375^{*}$ \cr 

    \hline 
    \multicolumn{4}{l}{$^{***}p<0.001$, $^{**}p<0.01$, $^*p<0.05$ }
  \end{tabular*}
  \caption{Nestedness coefficients and corresponding significance levels for the pooled and the sectoral R\&D networks in all observation years.
  The null hypothesis is that the measured nestedness coefficient is found in a random network with the same size and density as the giant component of the examined network: such a hypothesis can be rejected for most R\&D networks, especially during the ``golden age''.}
  \label{table:nestedness_pvalues}
\end{table}
\renewcommand{\arraystretch}{1.0}

We find that the null hypothesis cannot be rejected only for a few R\&D networks, especially outside of the ``golden age''.
However, during the ``golden age'', the p-values show that the null hypothesis can be rejected for all R\&D networks, confirming the presence of nested architectures that cannot be explained by taking into account only the network size and density, but are indicative of firms' peculiar alliance strategies.

\paragraph{Computation of the technological distance.}
The approach that we use to determine the knowledge position of a firm is to compute the shares of its patents in a set of different IPC classes.
The IPC, introduced in 1971 by the \textit{Strasbourg Agreement}, is a hierarchical system of symbols for the classification of patents according to the different areas of technology to which they pertain.\footnote{For more information on the International Patent Classification, see \url{http://www.wipo.int/classifications/ipc}.}
A generic IPC category consists of a letter, the so-called ``section symbol'', followed by two digits, the so-called ``class symbol'', and a final letter, the ``subclass''. This four-character term is then followed by a group/subgroup indication, represented by additional digits. A typical IPC term can be written as follows: B34H 6/99.
The sections identified by the IPC are historically stable and amount to 8, from A (human necessities) to H (electricity).
Given that we have to compute such a technological indicator on a broad set of firms, belonging to several industrial sectors,
 we have decided to consider only the section symbol (i.e. the first letter) in our empirical patent classification. 
Choosing a class- or subclass-level division would result in an excessive patent granularity,
\citep[see][for a more detailed discussion on the topic]{tomasello2015effect_on_technology}.
Next, we define the knowledge position of a firm $\mathbf{x}_i \equiv (x_{iA}, x_{iB}, \dots, x_{iH})$ as the set of normalized patent counts $x_{is}$ in each section: $x_{is} \equiv N_{is}/(\sum_s{N_{is}})$,
where $N_{is}$ is the number of patents that the firm $i$ has in a given IPC section $s$.
We then use the Euclidean metric, similar to \citet{tomasello2015effect_on_technology}, to compute the technological distance between two firms $i$ and $j$:
\begin{equation} \label{eq:patent_distance}
 |\mathbf{x}_i - \mathbf{x}_j| = \sqrt{\sum_{s=A}^{H}{(x_{is}-x_{js})^2}}.
\end{equation}
In particular, for the variable that we employ in our regression models, we consider only the patents for which the firm has applied in the last 5 years.
Note that, if the considered firm has no patent applications in this time window, its technological position is considered to be undetermined, thus generating a missing observation.

\bibliographystyle{apalike}


\end{document}